\documentclass[preprint,12pt]{elsarticle}

\usepackage{amssymb}
\usepackage{url}
\usepackage{lineno,hyperref}
\usepackage{amsmath,amssymb,amsfonts}
\usepackage{graphicx}
\usepackage{textcomp}
\usepackage{xcolor}
\usepackage{array}
\usepackage{url}
\usepackage{lineno,hyperref}
\usepackage{color,framed}
\usepackage{color,xcolor}
\usepackage{threeparttable}
\usepackage{multirow}
\usepackage{longtable}
\usepackage{ulem} 
\usepackage{setspace}
\usepackage{array}
\usepackage{booktabs}
\usepackage{amsmath}
\usepackage{amssymb}
\usepackage{lineno,hyperref}
\usepackage{listings,xcolor}
\usepackage{geometry}
\usepackage{courier}
\usepackage{lineno}
\usepackage{caption}   
\usepackage{graphicx}  
\usepackage{caption}
\usepackage{subfigure} 
\usepackage{pdflscape}
\usepackage{bbding}
\usepackage{adjustbox}
\usepackage{enumerate}
\usepackage{blkarray}
\usepackage{comment}
\usepackage{blkarray}
\usepackage{algpseudocode,float}
\usepackage{lipsum}
\usepackage[ruled,linesnumbered]{algorithm2e}
\SetKwRepeat{Do}{do}{while}

\newcommand{\reffig}[1]{Fig.\ref{#1}}
\newcommand{\refeqs}[1]{Eq.\ref{#1}}

\makeatother

\journal{Arxiv}

\begin{document}

\begin{frontmatter}



\title{An Efficient Dynamic Transaction Storage Mechanism for Sustainable High Throughput Bitcoin}


\author[inst1]{Xiongfei Zhao}
\ead{yb97480@um.edu.mo}

\affiliation[inst1]{organization={University of Macau},
            addressline={Department of Computer and Information Science}, 
            city={Macau}}
\fntext[myfootnote]{Corresponding author}

\author[inst1]{Gerui Zhang\fnref{myfootnote}}
\ead{mc15472@um.edu.mo}

\author[inst1]{Yain-Whar Si\fnref{myfootnote}}
\ead{fstasp@umac.mo}

\begin{abstract}
As coin-based rewards dwindle, transaction fees play an important role as mining incentives in Bitcoin. In this paper, we propose a novel mechanism called Efficient Dynamic Transaction Storage (EDTS) for dynamically allocating transactions among blocks to achieve efficient storage utilization. By leveraging a combination of Cuckoo Filter and Dynamic Transaction Storage (DTS) strategies, EDTS is able to improve the scalability while remaining sustainable even after the Bitcoin enters a transaction-fee regime. In addition to preventing deviant mining behaviors under the transaction-fee regime, EDTS can also provide differentiated transmission priorities based on transaction fees while allowing the investors to engage in pledging more transaction fees. In EDTS, we applied the multi-objective optimization algorithm U-NSGA-\uppercase\expandafter{\romannumeral3} to find the best DTS strategy and its corresponding attributes. Experimental results show that the EDTS mechanism together with the optimized DTS strategy can achieve a throughput of \textcolor{black}{325.3} TPS. The experimental results reveal that the scalability improvement of EDTS is superior to the performance of Bitcoin NG, which is the best known on-chain scaling solution, while maintaining the sustainability under the transaction-fee regime.
\end{abstract}




\end{frontmatter}


\section{Introduction}

Blockchains such as Bitcoin \cite{nakamoto2008Bitcoin}, Ethereum \cite{ethereum.org}, and Litecoin \cite{litecoin} are considered to be an ideal carrier of value transfer systems due to their robustness, credibility, transparency and other advantages. To ensure that no one can monopolize more than 51\% of the computing power to launch an attack like the double spending \cite{10.1145/2382196.2382292}, Blockchain mining incentives play a vital role as an economic incentive. It is also designed to ensure that hashing power is distributed among a sufficient number of different miners. 

However, as the major component of mining incentive, the supply of Bitcoin is capped at 21 million, and the \textcolor{black}{Bitcoin} mining reward is preset to be halved after mining for each group of 210,000 blocks. Starting from 50 BTC, mining reward for each block halves approximately every four years and will gradually be replaced by transaction fees. As coin-based rewards dwindle, transaction fees play an important role as mining incentives, rather than merely acting as an incentive for miners to include the transactions in their blocks.

Moreover, although transaction fees are bound to have a profound impact on block rewards when the \textcolor{black}{Bitcoin} evolves, the impact is already unfolding before the \textcolor{black}{Bitcoin} transits into transaction-fee regime. When users have a desire to settle transactions early, by paying a higher transaction fee, their transactions can get into blocks faster and, hence, they are able to settle their transactions faster. The bottleneck of the transaction processing capacity of \textcolor{black}{Bitcoin} directly leads to the increase of transaction fees. As transaction volume continue to rise, transaction fee have become increasingly important as part of the \textcolor{black}{Bitcoin}'s reward mechanism.

Since the security of the \textcolor{black}{Bitcoin} consensus protocol depends on the correct behavior of miners, the reward structure of the consensus protocol should encourage honest miners by ensuring that their earnings are proportional to their mining. However, due to the joint influence of the \textcolor{black}{Bitcoin}'s fixed block rewards halving and limited transaction processing capacity, the proportion of transaction fees continuously increase in block rewards. As a result, there is a growing possibility that miners' rewards could mismatch with their computing power. Profit-seeking miners may engage in deviant behaviors that could threatens the security and integrity of the \textcolor{black}{Bitcoin} network.

\begin{figure}[htp]
    \makeatletter
    \def\@captype{figure}
    \makeatother
    \centering
    \includegraphics[width=0.9 \textwidth]{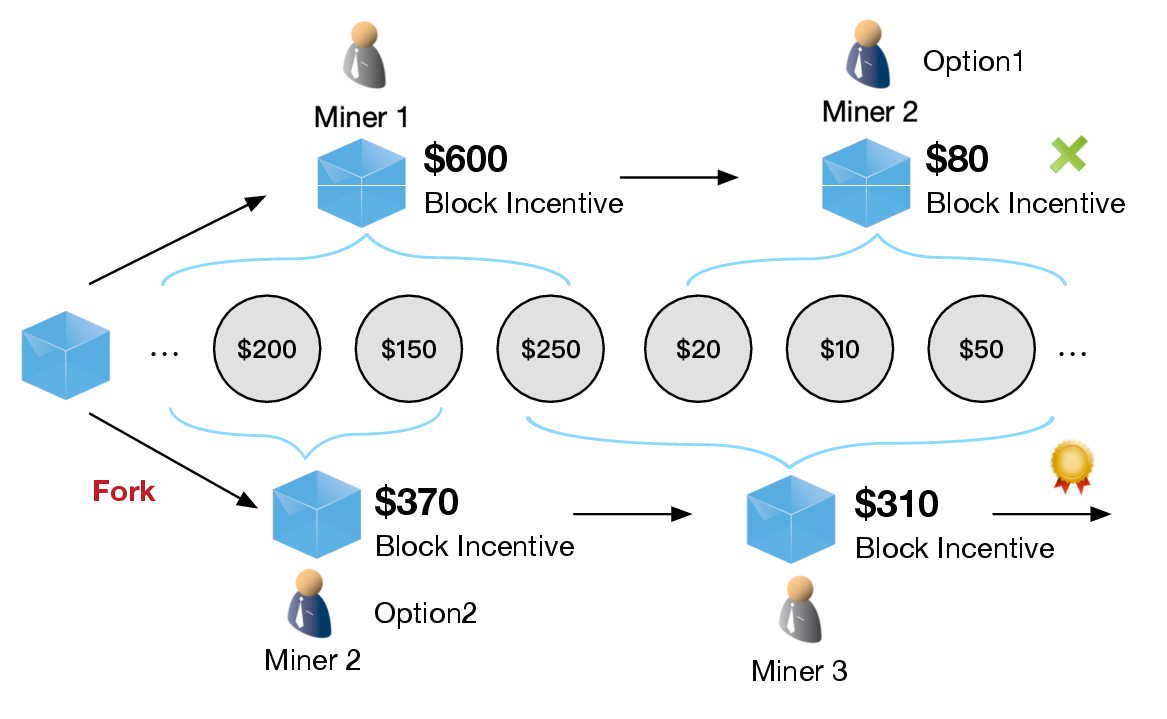}
    \caption{Undercutting Strategic Deviation}
    \label{UnderCutting}
\end{figure}

In \cite{10.1145/2976749.2978408}, Carlsten et al. proposed a deviant behavior of miners called Undercutting. \reffig{UnderCutting} illustrates how Undercutting works. After miner 1 mined a new block, it is miner 2’s turn to extend the longest chain. Miner 2 has two options. In option 1, miner 2 can obtain a reward of \$80. In option 2, miner 2 could fork the chain by obtaining a reward of \$370 and leave \$250 unclaimed for the next miner. As a result, option 2 will increase miner 2’s reward. Meanwhile, miner 2’s forking has a better chance to become the longest chain. By leaving unclaimed transactions, attackers can easily incentivize more miners to support his fork instead of the longest chain.
    
The limit in block size and the difficulty in adjusting the settings in conjunction with the PoW consensus protocol are some of the reasons behind the bottleneck of Bitcoin transaction processing capacity. Bitcoin processes seven transactions per second because it can only generates a 1MB block every 10 minutes. Users typically need to wait more than one hour (six blocks) to achieve a high level of assurance of the final state of a transaction. 
To solve the transaction processing capacity problem, various solutions have been proposed. These solutions can be classified into two categories, namely on-chain scaling and off-chain scaling. On-chain solution refers to the methods that increase the scalability by modifying only elements within the Bitcoin. These solutions include Bitcoin NG \cite{eyal2016Bitcoin}, Merkelized Abstract Syntax Tree (MAST) \cite{rubin2014merkelized}, Segregated Witness (Segwit) \cite{lombrozo2015segregated}, etc. The aim of the off-chain solution is to improve the scalability by processing the transactions outside of the Blockchain. Typical examples include Bitcoin's Lightning network \cite{poon2015Bitcoin} and Sharding \cite{luu2016secure}.

During the process of transition to a transaction-fee regime, Bitcoin's transaction processing capacity scaling proposal could be able to alleviate the contradiction between fast settlement and transaction fees. However, without a steady block reward as a guarantee, misbehaving miners can still seek to enhance their chance to fork the chain by giving up part of the available transaction fees. To the best of authors' knowledge, none of the scaling proposal can prevent such deviant miner behavior as transaction fee plays an important role when the coin-based reward dwindles. Although proposals such as Segwit and Bitcoin Cash allow more transactions to be incorporated into one block, these solutions can lead to greater imbalance of block rewards when there is a spike in transaction fee  due to network congestion.

In the authors' previous work \cite{9592512}, the Dynamic Transaction Storage (DTS) strategies were proposed to counter the halving of Bitcoin fixed block rewards. The core concept of DTS is to utilize the fixed block size to dynamically allocate transactions based on transaction fees to maintain stable mining rewards. DTS strategies can convert time-varying transaction fees into stable block rewards. In general, DTS can be applied to any Blockchains which are under transaction-fee regime. DTS can also coexist with existing coin-based rewards to stabilize block incentives. 

Based on the concept of DTS, in this paper, we propose a novel mechanism called Efficient Dynamic Transaction Storage (EDTS) which is designed to improve both scalability and sustainability of Blockchains which are under transaction fee regime. In the proposed EDTS mechanism, transactions that are supposed to be incorporated into a block are fed into a Cuckoo Filter. The block body which is originally used to store the transaction details, is  modified to store only one Coinbase transaction and a number of inspector transactions. By doing so, EDTS allows more transactions to be fitted into a block without changing the block size. To maintain the volatility of mining reward within a low range, the proposed EDTS mechanism also adopts DTS strategies \cite{9592512} which use several key attributes to control how the transactions are dynamically incorporated into a block. Besides, multi-objective optimization algorithm U-NSGA-\uppercase\expandafter{\romannumeral3} is also used to find the best DTS strategy and its corresponding attributes. Based on the data compression obtained by the Cuckoo Filter and by using the optimized DTS strategy, the proposed EDTS mechanism is able to achieve the properties such as efficient dynamic transaction allocation and block propagation. Experimental results reveal that the EDTS mechanism combined with the optimized DTS strategy can achieve a throughput of \textcolor{black}{325.3} TPS. The contributions of our proposal are three folds:

\begin{itemize}

    \item[a)] \textcolor{black}{Bitcoin based on proposed EDTS is capable of achieving up to \textcolor{black}{325.3} TPS, which is over 45 times the throughput of current Bitcoin \cite{nakamoto2008Bitcoin}. Besides, the scalability performance of the EDTS mechanism is better than the best known Bitcoin on-chain scaling solution \cite{eyal2016Bitcoin} \cite{9133427}.}
    
    \item[b)] In addition to effectively improving Bitcoin's throughput, EDTS have a unique advantage in keeping Bitcoin sustainable. By comparing the historical mining rewards volatility range over a 10-year period, EDTS is proven to be able to maintain block reward stable even after Bitcoin is operating under transaction-fee regime.
    
    \item[c)] U-NSGA-\uppercase\expandafter{\romannumeral3} \cite{seada2015u} multi-objective optimization algorithm is used to optimize the attributes of DTS strategies. By combining the EDTS mechanism with the optimized DTS strategies, we are able to achieve high transaction processing capacity while maintaining the volatility in an acceptable range. The results obtained in this study can be directly applied to Bitcoin without further optimization of DTS strategies.
    
\end{itemize}

\textcolor{black}{This paper is organized as follows. Section 2 summarizes the related work. Section 3 introduces the Preliminaries. In section 4, we describe the framework for Efficient Dynamic Transaction Storage (EDTS) mechanism. Section 5 describes the experiment settings and experimental results. Section 6 evaluates the results in terms of scalability, sustainability and prioritization. The paper is concluded in Section 6.}

\section{Related Work}

\textcolor{black}{After the coin-based rewards diminish gradually, transaction fees are expected to become a major mining incentive to maintain the sustainability of Blockchains. In \cite{10.1145/2976749.2978408}, Carlsten et al. reported about potential issues when Bitcoin mining transitions into a transaction-fee regime. The time-varying nature of transaction fees can lead to a richer set of strategic deviations such as \textsl{Selfish Mining}, \textsl{Undercutting}, \textsl{Mining Gap}, \textsl{Pool Hopping}.}

\textcolor{black}{In~\cite{basu2019stablefees}, Basu et al. proposed StableFees which is a second-price-auction based fee settling mechanism. In StableFees, users only need to pay their own transaction fees based on the lowest fee of all transactions in the block, and miners will receive the average revenue generated by the most recently mined blocks. Also in~\cite{10.1145/3479722.3480991}, Matheus et al. proposed a dynamic posted-price mechanism. When it is compared to the first-price auction, the proposed dynamic posted-price mechanism not only uses the block utilization, but also computes a posted-price for subsequent blocks using the observable bids from the past blocks. A monopolistic auction mechanism was also proposed by Ron et al. \cite{10.1145/3308558.3313454} in which all transactions in the block pay the identical fee and miners are able to choose a set of transactions that is likely to maximize their block incentive, thus decoupling the correlation between decreasing the block rewards and block size cap.}

\textcolor{black}{EIP1559 \cite{EIP1559}, which is also known as ``London Hard Fork", was launched on Ethereum on August 5, 2021. EIP1559 includes a base fee representing the minimum gas price a user needs to pay in each block. EIP1559 dynamically adapts and burns based on the parent block to reduce the impact of transaction fees on block incentives. EIP1559 allows users to specify the transaction fee bids. Users can set a priority fee per gas to incentivize miners to prioritize their transactions. User can also set a maximum fee per gas which is used to bound their total costs, including base fees as well as the priority fees. EIP1559 also introduces a variable and it can dynamically adjust the block size by 2x (from 15M to 30M) \cite{leonardos2021dynamical}. In EIP1559, the number of transactions each block can contain is largely determined by the base fee. One of the characteristics of EIP1559 is that the base fee is always burned and the miner can only keep the priority fee. Therefore, it removes the incentive for the miners to manipulate the fee.}

\textcolor{black}{Existing researches also target potential solutions on providing manipulation-proof and more predictable transaction fee mechanisms to prevent misbehavior from users and miners. However, their focus is from the perspective of a single transaction within a block, rather than the overall incentive fluctuation of different blocks in the Blockchain. Whereas in EIP1559's approach, it can also significantly affect the Fee-based rewards which are generated by transactions \cite{leonardos2021dynamical}. Therefore, deviant mining behaviors such as \textsl{Selfish Mining}, \textsl{Undercutting}, \textsl{Mining Gap}, and \textsl{Pool Hopping} \cite{10.1145/2976749.2978408} still cannot be solved by existing researches. Although the on-chain scalability improvement proposals such as Bitcoin NG, MAST, Segwit, BCC are feasible under coin-based mining rewards, none of them can adequately address the block incentive challenges in the future.}

As the number of users grows, to facilitate the increasing volume of transactions, the Bitcoin network needs to provide credibility to handle increasing number of transactions in the future. Various proposals have been introduced on how to scale Bitcoin. These approaches can be categorized as on-chain scaling and off-chain scaling. Since EDTS can be classified as on-chain solutions, in this paper, we mainly focus on on-chain solutions.

\begin{itemize} 

    \item \textbf{\textsl{Bitcoin NG \cite{eyal2016Bitcoin}}} Bitcoin NG divides the traditional block into two parts: key block and microblock, where the key block is used to elect the leader, and microblock is used to store transactions. To become a leader, miners need to compete with each other. Leaders are responsible for the generation of microblock until a new leader is chosen. This mechanism improves throughput and reduces latency.

    \item \textbf{\textsl{Merkelized Abstract Syntax Tree (MAST) \cite{rubin2014merkelized}}} MAST is the combination of a Merkle Tree and an Abstract Syntax Tree. MAST manage to fragment the programmining of the Bitcoin Script. By using Merkle trees to reduce the size of transactions, the space utilization of Blockchain has been improved. Without including all the information in the chain, MAST also improve the scalability issues that Blockchain network presents and reserved more space for processing more transactions. In addition, MAST also raises the transaction privacy level of the Blockchain by revealing only basic information about a transaction or script.

    \item \textbf{\textsl{Segregated Witness (Segwit) \cite{lombrozo2015segregated}}} Since the storage space required for digital signatures accounts for more than 70\% of the transaction information, the data structure named ``Witness” is adopted in Segwit, which separates the signature information from the transaction. The data capacity after removing certain transaction information can be used to bundle additional transactions per second. Segwit can handle 1.7 to 4 times more transactions than before since it solves the quadratic hashing problem and thus reducing transaction costs.

    \item \textbf{\textsl{Bitcoin Cash (BCC) \cite{Bitcoincash}}} Bitcoin Cash (BCC) is a peer-to-peer electronic cash system which is created from a hard fork of Bitcoin. In BCC, the block size is increased from 1 MB to 8 MB. Later, the block size is increased to 32 MB to handle more transactions. In BCC, while blocks are still generated at a 10-minute interval, throughput is much higher and transaction fees are reduced. 

\end{itemize} 

In comparison, our proposed efficient Dynamic Transaction Storage (EDTS) mechanism provides a novel solution that not only improves the scalability of the Blockchain, but also could stabilize incentive fluctuation of different blocks in the Blockchain.

\section{Preliminaries}

In the following section introduces recent known threats on stability due to the increasing proportion of transaction fees in the block rewards. Next, we introduce DTS strategies that could prevent such instability threats. Finally, we introduce the Cuckoo Filter algorithm, which plays a crucial role in our proposed EDTS mechanism.

\subsection{Instability of Bitcoin}

With the decline in the numbers of Bitcoin, transaction-fee regime is poised to play an increasingly important role. After coin-based rewards reach zero, the transaction-fee regime is expected to become a major mining incentive resource to maintain the sustainability of the Blockchain. 
Besides, the time-varying nature of transaction fees allows a richer set of strategic deviations such as \textsl{Selfish Mining}, \textsl{Undercutting}, \textsl{Mining Gap}, and \textsl{Pool Hopping}.

\begin{itemize} 

    \item \textbf{\textsl{Selfish Mining}} High volatility in mining incentives could lead to miner's misbehavior. For example, selfish miners could hold newly mined blocks without disclosing them. Such behavior enables the miners to get more than their fair share of rewards \cite{10.1007/978-3-662-45472-5_28}. Also, in a transaction-fee regime, selfish mining is immediately profitable \cite{10.1145/2976749.2978408}.

    \item \textbf{\textsl{Undercutting}} \cite{10.1145/2976749.2978408} According to Bitcoin mining protocol, each miner can decide what and how many transactions to include in their block. That also means miner can deliberately leave high-fee transactions in the mempool to attract other miners to extend their chain. If there is a 1-block fork, it is more profitable for the next miner to break the tie by extending the block that leaves the most available transaction fees rather than the oldest-seen block.

    \item \textbf{\textsl{Mining Gap}} \cite{10.1145/2976749.2978408}  Without a block reward, immediately after a block is found, there is zero expected reward for mining but nonzero electricity cost. Uncertainty in mining rewards, and even unprofitable risk, makes it difficult for the miners to estimate their return on investment (ROI) of mining power and reduce their incentive to participate in mining activities.

    \item \textbf{\textsl{Pool Hopping}} \cite{DBLP:journals/corr/abs-1112-4980} \textcolor{black}{A mining pool is formed when a group of miners combine their computational resources into a pool to improve the probability of finding a block. Miner's expected reward for participation varies over time, depending on how many shares have been contributed since the pool found its last block. In pool hopping, miners would respond by ``hopping" in real-time to the pool that maximizes their expected rewards.}

\end{itemize} 

\subsection{Dynamic Transaction Storage (DTS) Strategies}

\begin{figure}[htp]
    \makeatletter
    \def\@captype{figure}
    \makeatother
    \centering
    \includegraphics[width=0.7 \textwidth]{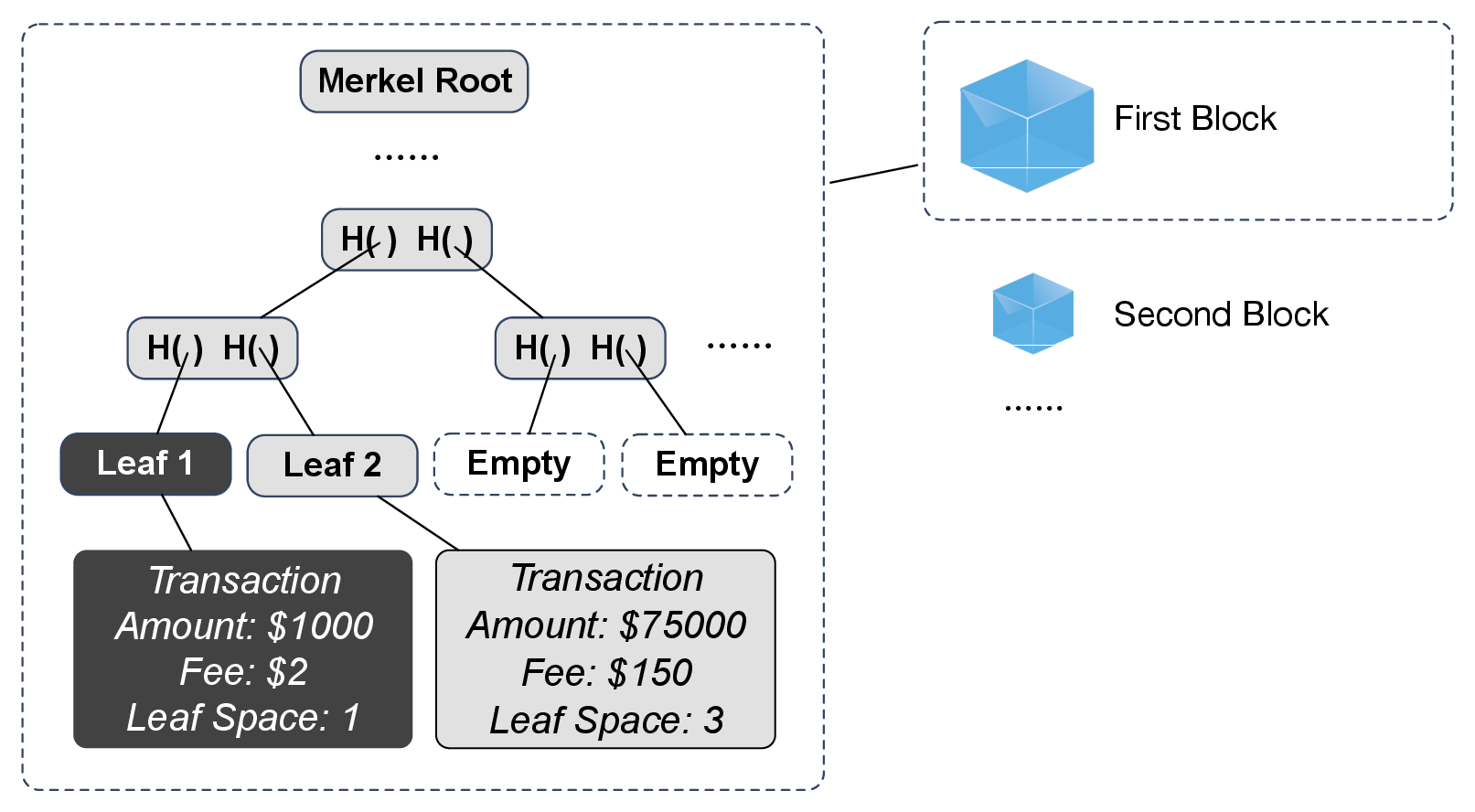}
    \caption{Illustration of the Dynamic Transaction Storage (DTS) strategies transaction
incorporation.}
    \label{DTS}
\end{figure}

In \cite{9592512}, Zhao et al. proposed a set of Dynamic Transaction Storage (DTS) strategies for Bitcoin. \textcolor{black}{In these strategies, the process of incorporating the transactions in a block was extended. That is,  transactions are incorporated in blocks more flexibly to harmonize block incentive while maintaining consensus protocols.} As shown in \reffig{DTS}, in DTS strategies, transactions are dynamically incorporated into the blocks based on each transaction's amount. By converting the time-varying transaction fees into stable block rewards in DTS approach, deviant mining strategies \cite{10.1145/2976749.2978408} that could harm the stability of Bitcoin can be prevented. Besides, DTS strategies in turn can positively influence the block transfers at the network level. Higher-fee transactions take up more space in the block, which results in a smaller sized block containing higher-fee transactions. For a block to be diffused through the Bitcoin network, blocks propagate in hops. Therefore, the reduction in the number of transactions lead to a smaller block size which in turn reduce the validation and transmission delay at each hop during the propagation \cite{6688704}. However, the reduction in the number of transactions can effect the throughput of the Bitcoin. To alleviate this problem, in this paper, we propose an approach to increase the throughput while maintaining the volatility of the transactions within an acceptable range.

DTS strategies \cite{9592512} dynamically control the number of transactions incorporated based on the transaction fees. For each incorporated transaction, Cumulative Distribution Function (CDF) \refeqs{con:cdf} dynamically allocates the leaf space $(LeafSpace)$ that a transaction $(T)$ can occupy.

{\begin{equation}
CDF(x) = Pr(X\leq x) = \frac{1}{2} \, + \, \frac{1}{2} \text{erf} \bigg[\frac{ln(x)-\mu}{\sigma\sqrt{2}} \bigg] \label{con:cdf}
\end{equation}}

Based on the transaction fee $(fee)$ and the maximum leaf space $(MaximumSpace)$, the leaf space of the transaction $(T)$ can be calculated as follows:

{\begin{equation}
LeafSpace(T) = CDF(fee) * MaximumSpace \label{leaf}
\end{equation}}

DTS strategies uses a set of attributes to control how transactions are dynamically incorporated into a block. The definitions of these attributes are listed as follows  \cite{9592512}:

\begin{itemize}
	
\item \textcolor{black}{\textbf{\textsl{A1. Mempool Size:}} Mempool size represents the size of the holding area where a node keeps all the pending transactions.}

\item \textcolor{black}{\textbf{\textsl{A2. Transaction Incorporation Priority:}} There are two types of transaction incorporation priority. The priority for transactions processed on a first-come-first-served basis is classified as Time-based. The priority of transactions processed on the basis of the highest fee is classified as Fee-based.}

\item \textcolor{black}{\textbf{\textsl{A3. Transaction Fee Percentage:}} Transaction fees are charged as a percentage of the transaction amount which is closely related to the urgency of the transaction.}

\item \textcolor{black}{\textbf{\textsl{A4. Designated Space for Small-fee Transactions:}} Certain block space is designated for all the transactions whose fees are below a certain threshold. This attribute is used to control the percentage of small transactions per block.}

\item \textcolor{black}{\textbf{\textsl{A5. Small-fee Transaction Fee Threshold:}} When A4 is set to true, this attribute determines whether the transaction is a small transaction or not based on the amount of the fee.}

\item \textcolor{black}{\textbf{\textsl{A6. Small-fee Transaction Number Threshold:}}} When A4 is set to true, this attribute determines how many small-fee transactions can be incorporated into a block.

\item \textcolor{black}{\textbf{\textsl{A7. Maximum Space for One Transaction:}} This attribute controls the maximum ``\textsl{Leaf}'' space of the Merkle Tree for each transaction. This attribute allows the transactions to be reasonably allocated.}

\item \textcolor{black}{\textbf{\textsl{A8. Scale:}} The ($\mu$) in \refeqs{con:cdf} is called the \textsl{Scale} parameter. This attribute determines the statistical dispersion or ``scale'' of the probability distribution.}
    
\item \textcolor{black}{\textbf{\textsl{A9. Shape:}} The ($\sigma$) in \refeqs{con:cdf} is called \textsl{Shape} parameter. this attribute is the standard deviation of the normally distributed natural logarithm.}

\end{itemize}

\subsection{Cuckoo Filter}

\textcolor{black}{Cuckoo Filter \cite{fan2014cuckoo} is a space-efficient data structure for approximating the set membership tests. It is a probabilistic method similar to the Bloom Filter. However, Cuckoo Filter has better lookup performance and supports dynamic deletion of items than Bloom Filter. 
The upper bound of space cost (bits) of the Cuckoo Filter is obtained by \refeqs{cuckoo_cost}, where $\epsilon$ is the false-negative rate (FPR), and $\alpha$ is the load factor. When $FPR>3\%$ the Cuckoo Filter saves more space in the same FPR than $1.44log_2(1/\epsilon)$ space cost of the the Bloom Filter.}

\textcolor{black}{A Cuckoo Filter contains many buckets. A bucket remains empty or stores a fingerprint hashed from the input. While inserting a new item, the hash of the item specifies the bucket to store the fingerprint. Querying whether an item exists is a simple process. The Cuckoo Filter computes the location of the item and checks if there is the same fingerprint.}

\begin{equation}
    C \leq [log_2(\frac{1}{\epsilon})+3]/\alpha \label{cuckoo_cost}
\end{equation}

\section{Overview of proposed Efficient Dynamic Transaction Storage (EDTS) mechanism}

In order to scale the Bitcoin network, a faster but less bandwidth intensive method is needed to send larger blocks. \textcolor{black}{To this end, we utilize the transactions that already exist in the nodes' memory pool to rebuild the blocks, rather than downloading the entire block from other nodes.}

This paper proposes a novel Efficient Dynamic Transaction Storage (EDTS) mechanism in which the miner who wins the mining competition generates a space-efficient block instead of a normal block. EDTS takes advantage of the block reward stability by adopting the Dynamic Transaction Storage (DTS) strategies. EDTS also constructs a space-efficient block by using the Cuckoo Filter. In EDTS approach, miners are rewarded by the sum of transaction fees for all the transactions in that Block. In the proposed mechanism, transaction-fee based rewards per block are stable since the transactions are incorporated into a block based on the DTS strategies. In summary, EDTS is the first solution to achieve the efficient block propagation in the network while keeping the Bitcoin sustainable.

\begin{figure}[htp]
    \makeatletter
    \def\@captype{figure}
    \makeatother
    \centering
    \includegraphics[width=1 \textwidth]{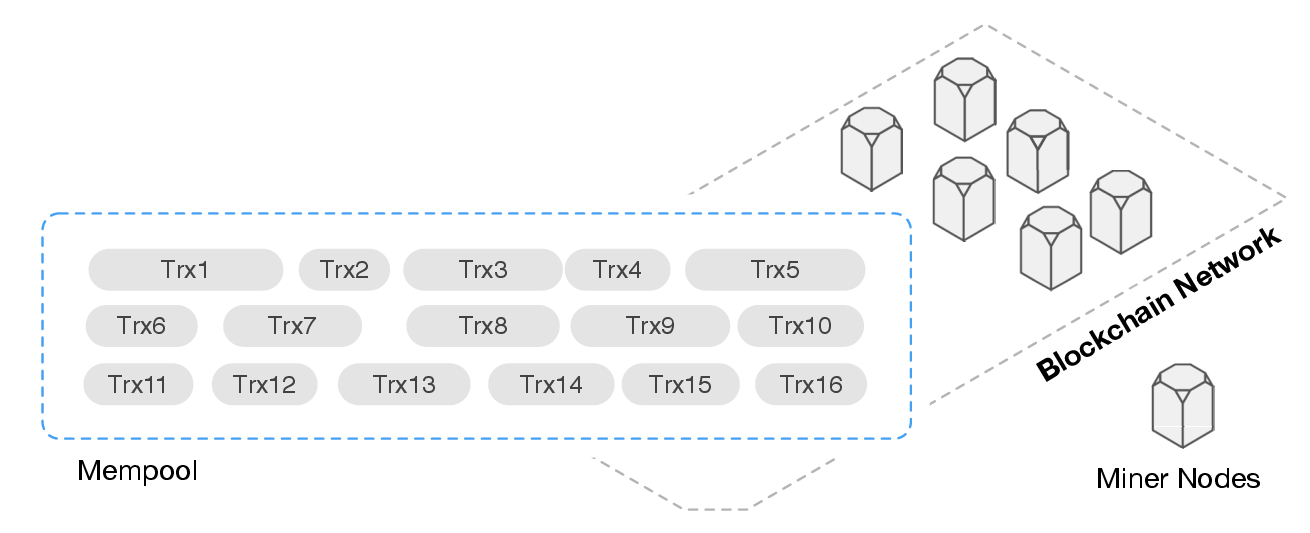}
    \caption{Illustration of the Blockchain network which is composed of nodes and an example mempool with unconfirmed transactions for each node.}
    \label{Mempool}
\end{figure}

In the current Bitcoin implementation, miners are incentivized to incorporate as many transactions as possible in the order of highest to lowest fees paid. To this end, Miners are connected to a large number of nodes to make sure that they get the complete transaction information. Mempool contains the set of valid transactions that the miner is aware of but has not yet been included in a block. Transactions in Bitcoin network are disseminated among nodes via flooding. Different miners has their own mempool, which contains valid transactions that have not yet been incorporated into the blocks (\reffig{Mempool}). Mempool size represents the size of holding area where the node keeps all the pending transactions. After a transaction is incorporated into a block, it is removed from the mempool. Transactions in different miners' mempools may vary due to the delay in transaction relay.

In the proposed EDTS mechanism, we make the following assumptions with respect to the current Bitcoin implementation:

\begin{itemize}
    \item \textcolor{black}{In this paper, for the purpose of simplification, we assume that miners have the same mempool.  Problems associated with inconsistent mempools for miner in Bitcoin should be treated separately. Since the EDTS mechanism utilize transactions in nodes' mempool to rebuild blocks, it is crucial for transactions of different mempool in EDTS can be synchronized in a timely manner. To archive efficient transaction relay between different nodes, solutions such as Erlay \cite{10.1145/3319535.3354237} has been proposed to minimize the number of redundant transactions transmitted by having peers only exchange sets of unsent transactions.} Erlay ensures that each node could receive each transaction only once, improving the efficiency of transaction synchronization and avoids the waste of network bandwidth resources.

    \item To evaluate the throughput of EDTS, we use the historical volatility range of block rewards of Bitcoin as a reference. We obtained total coinbase block rewards and transaction fees paid to the miners per day from the NASDAQ website. The records are within the range from 2012 to 2021 (a continuous period of 365 blocks for each year) \cite{nasdaq}. We denote the block reward from the records as $R$. Next, we downloaded the number of blocks $B$ mined on the Bitcoin network every day from 2012 to 2021 \cite{btc} from BTC.com. \textcolor{black}{By using the following formula, we can calculate average block reward on day $n$ denote as $A_{n}$ ($n=365$ for each year):}

    \begin{equation}
        \resizebox{.16\hsize}{!}{$A_{n}=R_{n}/B_{n}$}\label{daily}
    \end{equation}
    
    \textcolor{black} {First, we calculate the day-to-day returns $R$:}
    
    \begin{equation}
        \resizebox{.22\hsize}{!}{$R_{n}=ln(A_{n}/A_{n-1})$}\label{reward}
    \end{equation}
    
    \textcolor{black} {Next, we calculate the standard deviation of the returns $R_{n}$. The average of $R_{n}$ can be defined as follows:}
    
    \begin{equation}
        \resizebox{.22\hsize}{!}{$R_{avg}=\frac{\sum_{i=1}^nR_{i}}{n}$}\label{average}
    \end{equation}
    
    \textcolor{black} {Then, the standard deviation of block incentive $\sigma$, also known as block incentive volatility of each year, is calculated as follows:}
    
    \begin{equation}
        \resizebox{.32\hsize}{!}{$\sigma = \sqrt{\frac{\sum_{i=1}^n(R_{i}-R_{avg})^{2}}{n-1}}$}\label{standard deviation}
    \end{equation}
    
    Table~\ref{HistoricalVolatility} lists the historical volatility result of Bitcoin's average daily block rewards from 2012 to 2021, which are calculated by using \refeqs{daily} - \refeqs{standard deviation}. The volatility of the Bitcoin block reward fluctuates between the lowest volatility of 0.037647 in 2019, and the highest volatility of 0.238111 in 2012. In this paper, the minimum and maximum values from Table~\ref{HistoricalVolatility} are used as the reference volatility range in our experiments.
\end{itemize}

\linespread{0.9}
\begin{table}[htp]
  \footnotesize
  \centering
  \caption{\textcolor{black}{Historical Volatility of Block Rewards From 2012 to 2021}}
  \begin{threeparttable}
    \begin{tabular}{p{1.5cm}<{\centering}|p{4.5cm}<{\centering}|p{1.5cm}<{\centering}|p{4.5cm}<{\centering}}
      \hline\hline
      \specialrule{0.00em}{1pt}{1pt} 
      \textbf{Year} & \textbf{Historical Volatility} & \textbf{Year} & \textbf{Historical Volatility}\\
      \specialrule{0.00em}{1pt}{1pt} 
      \hline
      \specialrule{0.00em}{1pt}{1pt} 
      \textbf{2012} & \textbf{0.238111} & 2017 & 0.063965 \\
      \specialrule{0.00em}{1pt}{1pt} 
      \hline
      \specialrule{0.00em}{1pt}{1pt} 
      2013 & 0.200857  & 2018 & 0.045616 \\
      \specialrule{0.00em}{1pt}{1pt} 
      \hline
      \specialrule{0.00em}{1pt}{1pt} 
      2014 & 0.218010  & \textbf{2019} & \textbf{0.037647} \\
      \specialrule{0.00em}{1pt}{1pt}  
      \hline
      \specialrule{0.00em}{1pt}{1pt}  
      2015 & 0.180948 & 2020 & 0.059485 \\
      \specialrule{0.00em}{1pt}{1pt}  
      \hline
      \specialrule{0.00em}{1pt}{1pt}  
      2016 & 0.073051  & 2021 & 0.044932 \\
      \specialrule{0.00em}{1pt}{1pt}  
      \toprule
    \end{tabular}
  \end{threeparttable}
  \label{HistoricalVolatility}
\end{table}

In the next section, we introduce the current Bitcoin mining process and block generation. Next, we introduce the a new block generation mechanism by combining the data structure of Cuckoo Filter with Dynamic Transaction Storage (DTS) strategies.

\subsection{Existing transaction incorporation process}

\begin{figure}[htp]
    \makeatletter
    \def\@captype{figure}
    \makeatother
    \centering
    \includegraphics[width=0.65 \textwidth]{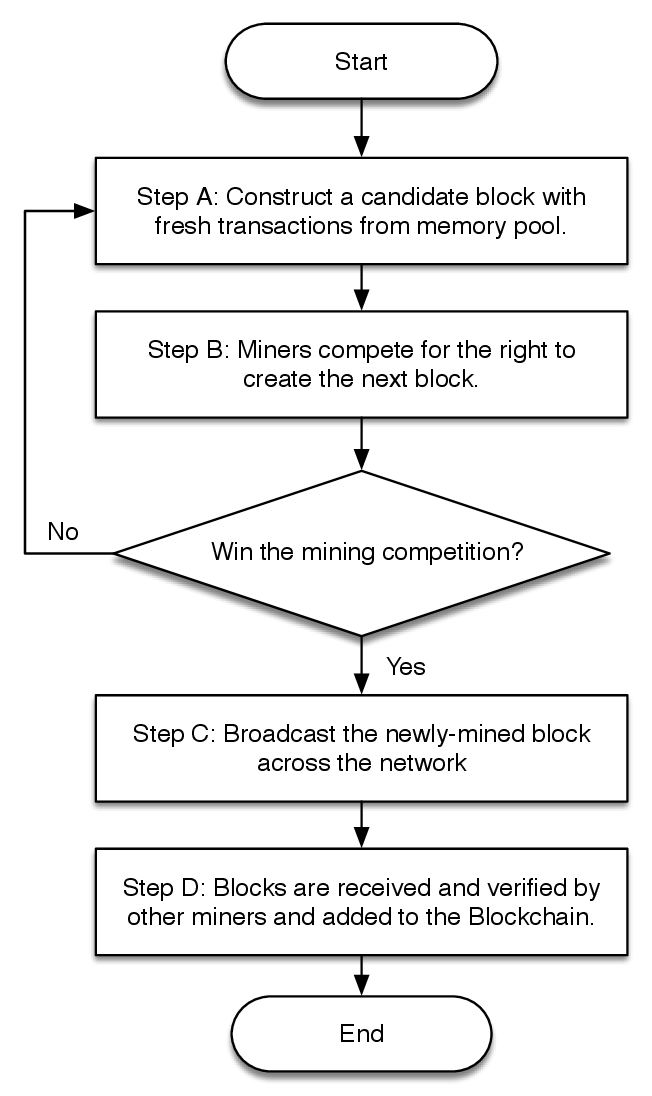}
    \caption{Existing Bitcoin network mining process}
    \label{Mining}
\end{figure}

As shown in \reffig{Mining}, miners nodes compete for the right to create the next block by using their own resources (i.e., computing power to solve a mining puzzle often denoted as proof of work). Since no one has a monopoly on all the available resources, all miners have equal opportunity to fight for the creation of blocks. When a miner solves the computational puzzle first, it has the right to insert a series of transactions proposed by its node into the block and receive a reward in the form of a Bitcoin (or other cryptocurrencies) for participating.

\begin{figure}[htp]
    \makeatletter
    \def\@captype{figure}
    \makeatother
    \centering
    \includegraphics[width=1 \textwidth]{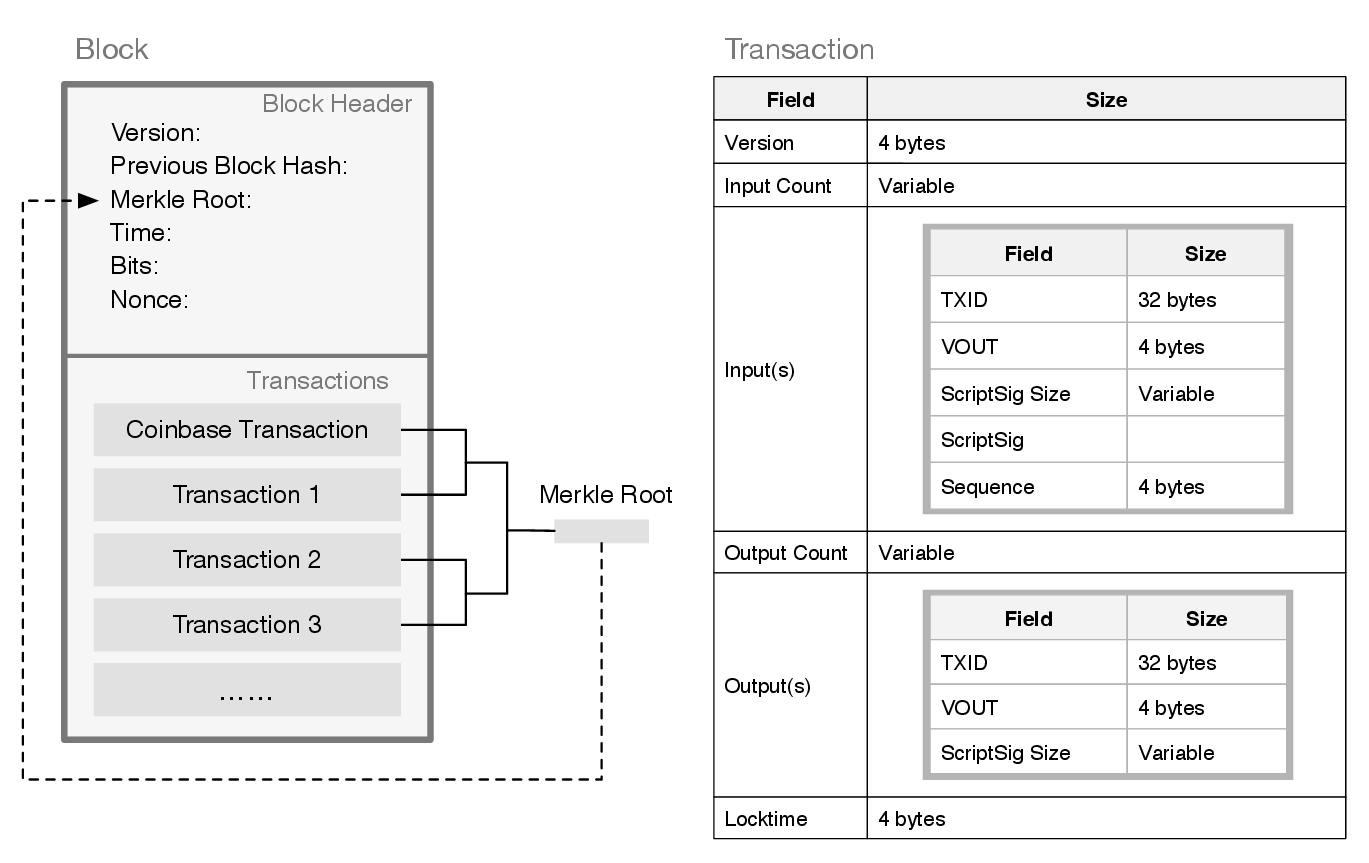}
    \caption{Block and transaction data structure of current Bitcoin network}
    \label{Block structure}
\end{figure}

The mining process begins by filling a candidate block with transactions from the miner node’s mempool. As shown in the left of \reffig{Block structure}, miner constructs a block header for this candidate block. This is basically a short summary of all the data inside the block, which includes a reference to an existing block in the Blockchain that miner wants to build on. \textcolor{black}{A Merkle Root is also created to provide a short and unique fingerprint of all transactions in a block. Merkel Root is created by hashing the transaction ID pairs. The transaction ID is denoted as TXID, which is the identification of a Bitcoin transaction obtained by hashing the transaction data twice through SHA256.}

A complete Bitcoin transaction structure is shown in the right part of \reffig{Block structure}, with version numbers (4 bytes), inputs, outputs, and lock time (4 bytes). The transaction input contains the reference transaction hash (32 bytes), the index number for an output in a transaction  (4 bytes), the input script information, and the serial number (4 bytes). The transaction output consists of the number of Bitcoins (8 bytes) and the output script information.

Each newly-mined block is broadcast across the network, where each node independently verifies it before adding it on to their Blockchain. After adding the new block, each node will stop working on their own candidate block, construct a new one (with fresh transactions from their memory pool), and start trying to build on top of this new block in the chain. As a result, miners are constantly working independently (yet collaboratively) to extend the Blockchain with new blocks of transactions. The Blockchain is constantly being built under a collaborative effort of nodes across the network

\subsection{Transaction incorporation based on proposed EDTS mechanism}
	
	In the proposed EDTS mechanism, we aim to improve the efficiency of block propagation by incorporating more transactions while maintaining the same block size. In EDTS, transaction details that are normally saved to the block body are omitted, and are instead added to the Cuckoo Filter. With EDTS, detailed transaction information are no longer mandatory for other miners to precisely generate the same block. When the other miners received the EDTS block, they can reconstruct the same block based on the detailed transaction information in their own mempool and the Cuckoo Filter in the block header. The reconstructed block also contains hash correlations from the original block and any future transactions referring to this reconstructed block can be validated. 

    \textcolor{black}{Next, we will introduce the detailed steps of integrating EDTS mechanism and DTS strategies. Under the Bitcoin model, transactions are organized in a Merkle tree structure, with each transaction recorded in the leaves of the tree, allowing people to digest and locate transactions in the block efficiently. First, based on the fix block size of 1MB, we calculate maximum number of transaction can be incorporated into a Cuckoo Filter within a desired false positive probability. Next, we utilize the DTS strategies \cite{9592512} to allocate transactions over a maximum number of Merkel Tree leaf nodes in a block dynamically.}

\begin{figure}[htp]
        \makeatletter
        \def\@captype{figure}
        \makeatother
        \centering
        \includegraphics[width=1 \textwidth]{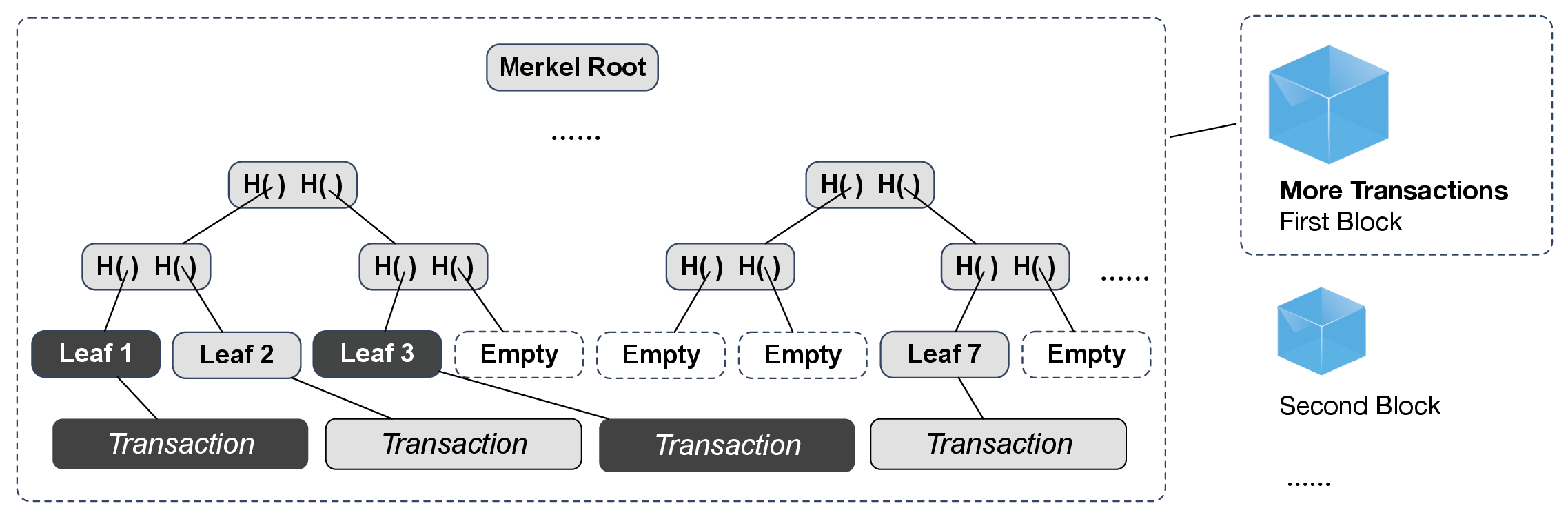}
        \caption{Illustration of the Efficient Dynamic Transaction Storage (EDTS) mechanism transaction incorporation.}
        \label{EDTS}
    \end{figure}

    \textcolor{black}{Through the EDTS mechanism, we can incorporate more transactions into one block while keeping the  block size unchanged. The larger block capacity can also compensate deficiencies of DTS mechanism in utilizing of block space. EDTS provides more block space for DTS strategies to dynamically allocate block space for different transactions according to the transaction fee. As illustrated in \reffig{EDTS}, more transactions are incorporated into the first block, while the second block is smaller because it incorporates transactions with higher fees.}
	
	For the purpose of simplification, in this paper, we denote the node that wins the mining competition and subsequently generates a new block as the ``sender". The other nodes that receive the propagated block as ``receivers". In the following sections, we introduce their working mechanism in detail.
	
	\subsubsection{Sender}
	
\begin{figure}[htp]
		\makeatletter
		\def\@captype{figure}
		\makeatother
		\centering
		\includegraphics[width=0.98 \textwidth]{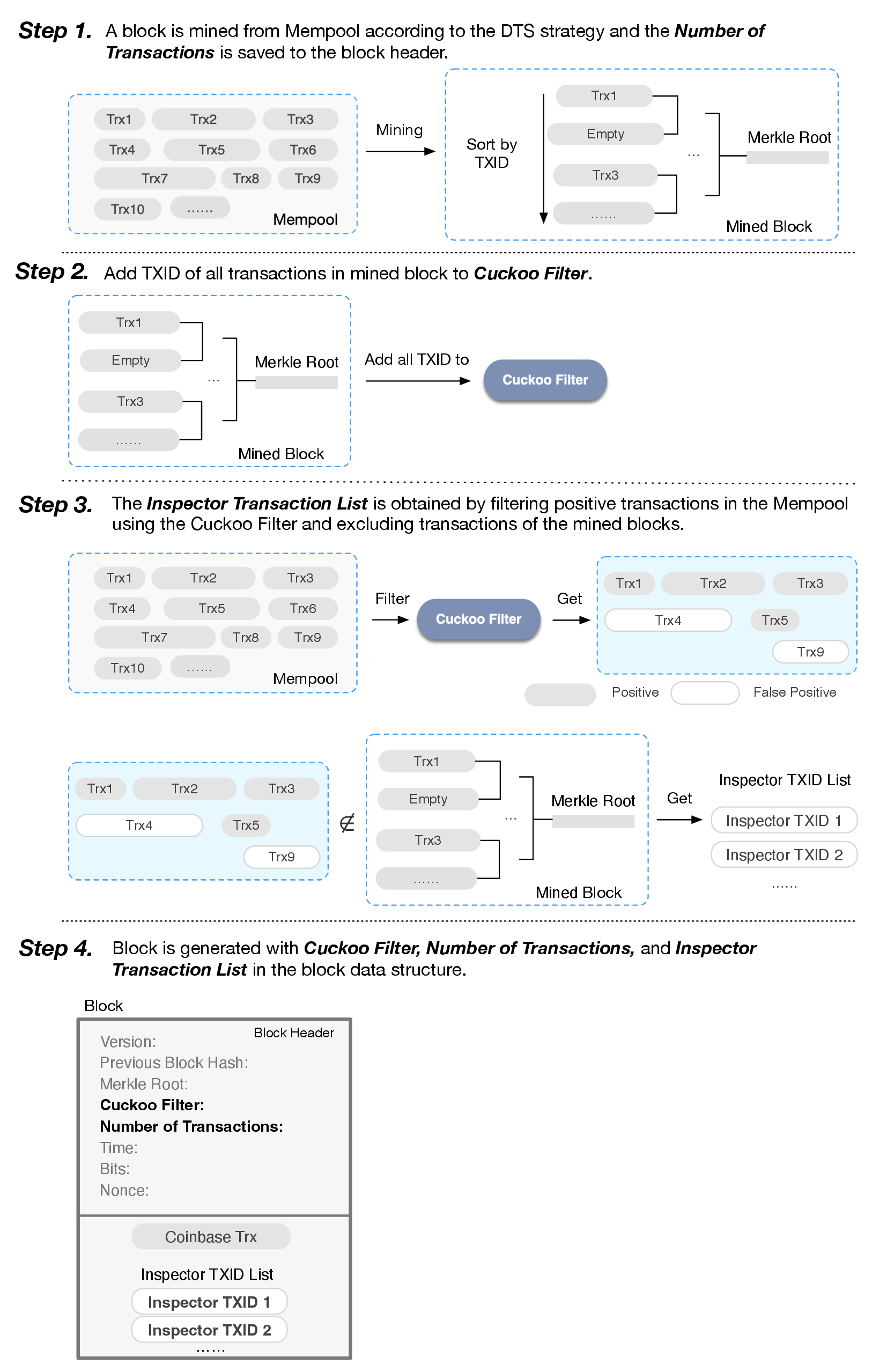}
		\caption{Schematic diagram of Sender's mining process based on EDTS mechanism}
		\label{MiningSteps}
	\end{figure}

	Schematic diagram of Sender's mining process under EDTS mechanism is depicted in \reffig{MiningSteps}. In this process, \textsl{``Cuckoo Filter''}, \textsl{``Number of Transactions''} and \textsl{``Inspector Transaction List''} are introduced to the block data structure when the miner constructs a candidate block. These newly introduced fields are derived from the following steps. These fields allow other miners to precisely construct the same blocks.
	
	\begin{itemize}
		\item \textbf{\textsl{Step 1.}} \textcolor{black}{When a miner wins the mining competition, it selects transactions from the mempool to be added to the mined block. EDTS adopts the DTS strategies to calculate the Merkle Tree leaf space that can be occupied for each transaction based on the transaction fees. The transactions incorporated into a mined block are sorted by their TXID. \textcolor{black}{The number of Transactions} in mined block will be saved to block header.}
		
		\item \textbf{\textsl{Step 2.}} \textcolor{black}{The transactions that are incorporated into the candidate block in Step 1 are added to the Cuckoo Filter and the total number of transactions is recorded in the block header. Next, each transaction's cuckoo hash fingerprints are mapped into two hash tables. The first table is for recording purpose and the second table is for handling collisions.}
		
		\item \textbf{\textsl{Step 3.}} The newly generated Cuckoo Filter is then used to detect whether a transaction belongs to the mined block. It also filters out the positive transactions from the mempool. Transactions that meet the criteria of the Cuckoo Filter are again compared with those transactions in the mined block. Eligible transactions that are not part of the mined block eventually form the ``Inspector Transaction List". \textcolor{black}{Due to the low collision probability of the Cuckoo Filter, the ``Inspector Transaction List" only contains a few transactions.}
		
		\item \textbf{\textsl{Step 4.}} Finally, the newly generated ``Cuckoo Filter" and the ``Number of Transactions'' are added to the block header. ``Inspector Transactions" are also stored in the block body. The block body also contains a special transaction that contains reward transaction for the miner, also known as as a Coinbase transaction. The Coinbase transaction includes a total transaction fee award for the current block.
	\end{itemize}
	
	Newly mined block is then propagated to the entire Bitcoin network using a Gossip protocol until the block reaches the entire network.  
	
	\newgeometry{left = 2.8 cm, top=0.8cm}
	
	\begin{figure}[htp]
		\makeatletter
		\def\@captype{figure}
		\makeatother
		\centering
		\includegraphics[width=0.9 \textwidth]{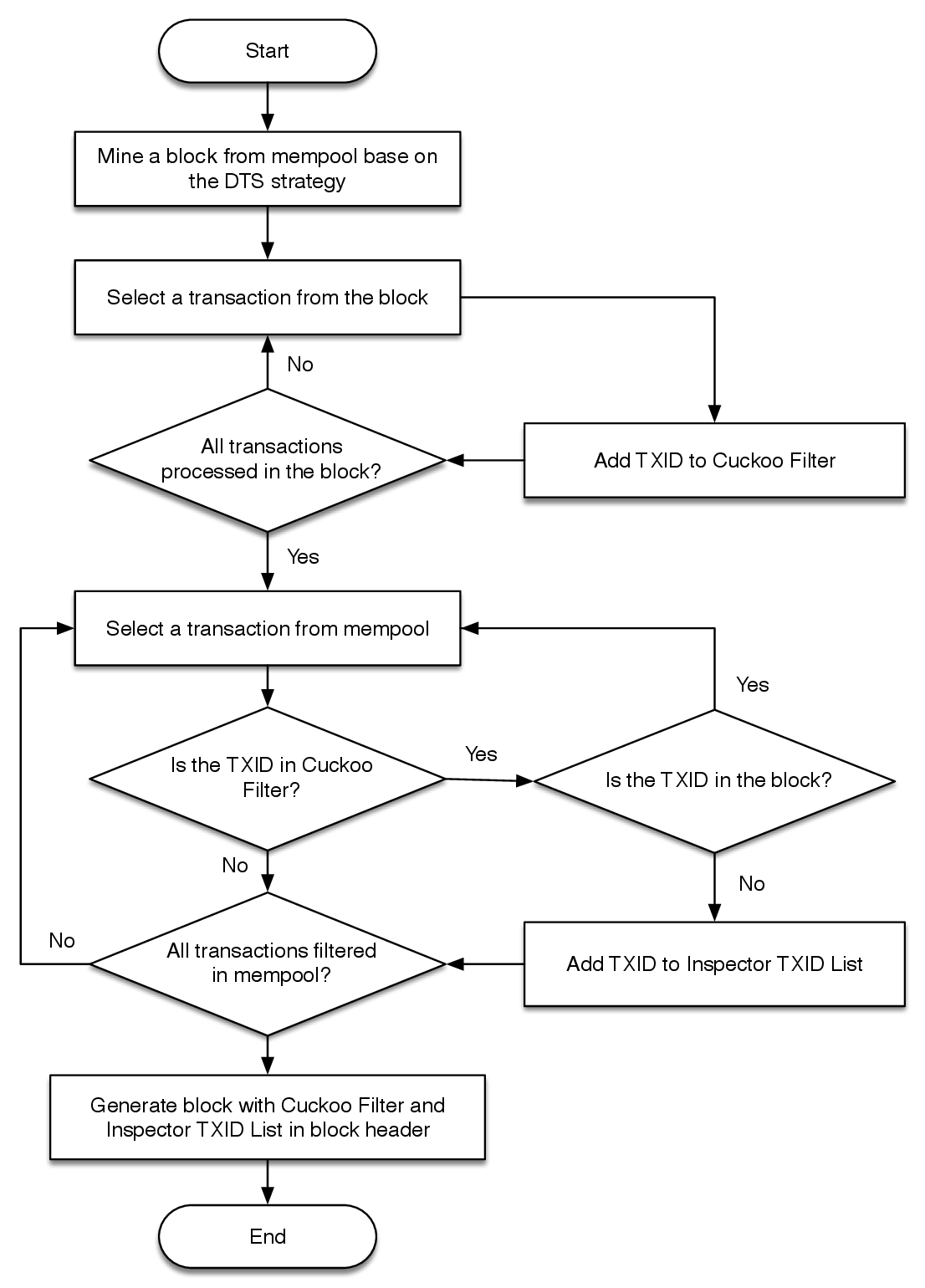}
		\caption{Flow chart of Sender's mining process under EDTS mechanism}
		\label{BlockGeneration}
	\end{figure}
	\restoregeometry
	
	\begin{algorithm}[H] 
		\footnotesize
		\setstretch{1.3}
		\caption{Sender's Block Mining Process Under EDTS Mechanism} 
		\SetAlgoLined  
		\KwData{Mempool $M$, Cuckoo Filter $C$, Cuckoo Filter's false positive rate $\epsilon$, Cuckoo Filter's hash table load factor $\alpha$, Cuckoo Filter's upper bound of space cost $B$, Leaf space for one block $L$, Transaction $T$, Number of transactions in the block $N$}  
		\KwResult{Efficient block $B$}
		\BlankLine 
		\tcp{initialization} 
		$\epsilon \leftarrow 10^{-12}$\; 
		$\alpha \leftarrow 0.955$\; 
		$C \leftarrow \frac{log_2(\frac{1}{\epsilon})+3}{\alpha*8}$\; 
		$L \leftarrow \frac{1024*1024-80}{C}$\;
		$N \leftarrow  0$\;
		\Do{$L > 0$}{
			Select $T$ from $M$\;
			Add $T$ to $C$\;
			\tcp{Cumulative Distribution Function CDF()}
			$L = L - CDF(T.Fee)$\;
			Incorporate $T$ into $B$\;
			$N \leftarrow N + 1$\;
		} 
		Add $C$ to $B$'s header as ``Cuckoo Filter''\;
		Add $N$ to $B$'s header as ``Number of Transactions''\;
		\ForEach{$T$ in $M$}{
			\If{$T \in C$}
			{
				\If{$T \notin B$}
				{
					Add $T$ to $B$'s body as ``Inspector Transaction''\;
				}
			}
		}  
	\end{algorithm}
	
	\subsubsection{Receiver}
	
\begin{figure}[htp]
		\makeatletter
		\def\@captype{figure}
		\makeatother
		\centering
		\includegraphics[width=0.98 \textwidth]{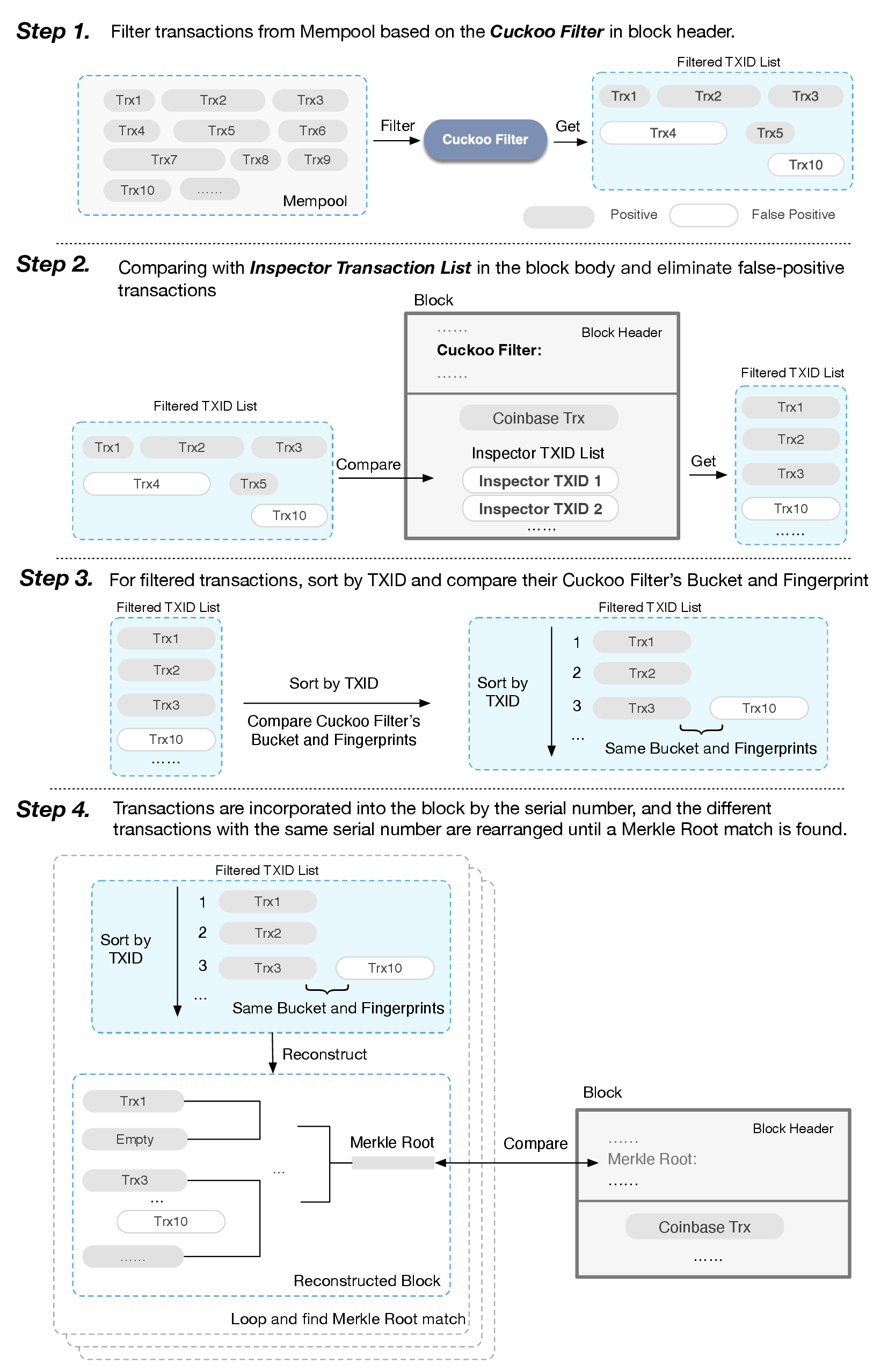}
		\caption{Schematic diagram of Receiver's mining process in the EDTS mechanism}
		\label{ReceivingSteps}
	\end{figure}

	Schematic diagram of Receiver's mining process under EDTS mechanism is depicted in \reffig{ReceivingSteps}. In this process, after receiving the block generated in EDTS mining process, miner will restore the transactions details based on its own mempool. The following steps describe filtering process using the Cuckoo Filter. The process identifies the required transactions with minimal computational complexity.
	
	\begin{itemize}
		\item \textbf{\textsl{Step 1.}} When a node receives a block, it first uses the Cuckoo Filter information in the block header to filter the transactions from its own mempool. \textcolor{black}{Filtered transactions form a Filtered TXID List and they are sorted by TXIDs.} As shown in \reffig{ReceivingSteps} Step 1, transactions Trx1, Trx2, Trx3, Trx4 etc. are matched by the Cuckoo Filter. However, Trx4 and Trx10 are false positive matches.
		
		\item \textbf{\textsl{Step 2.}} \textcolor{black}{Transactions in the Filtered TXID List are further verified by using the ``Inspector Transaction List" in the block body. As shown in \reffig{ReceivingSteps} Step 2, the list of ``Inspector Transactions" in the block body ensures that false-positive transactions (e.g. Trx4) that also exists in Miner's mempool are eliminated from the Filtered TXID List. }
		
		\item \textbf{\textsl{Step 3.}} \textcolor{black}{In general, if the items tested with a Cuckoo Filter generate false positive results, their buckets and fingerprints could have collisions with other non-false positive items \cite{fan2014cuckoo}. In EDTS, all transactions that generate the same bucket and fingerprint are recorded in Step 1. Assume that there are $m$ transactions from the Filtered TXID List whose buckets and fingerprints appear more than once. Also assume that $k$ transactions whose buckets and fingerprints only appear once. If the total number of transactions in the block $N$ is recorded in the block header, we only need to select a subset of $N-k$ transactions from $m$ transactions to combine with the remaining $k$ transactions. \textcolor{black}{A block is reconstructed and the Merkle root is calculated. The total number of iterations need to be performed is $C^{N-k}_m$ until the same Merkle root hash is found.}}
		
		\item \textbf{\textsl{Step 4.}} The transaction combination selected in the Filtered TXID List is finalized when the reconstructed block's Merkle Root is the same as the Merkle Root recorded in the block header. \textcolor{black}{Under some extreme scenarios, there will be a transaction in the receiver's mempool that passes the checking of Step 1 to Step 4, and the reconstructed block will have the same valid Merkle Root. In this case, the receiver can send these transactions back to the sender to verify which transactions are valid.}
	\end{itemize}
	
	\newgeometry{left = 2.8 cm, top=0.8cm}
	
	\begin{figure}[htp]
		\makeatletter
		\def\@captype{figure}
		\makeatother
		\centering
		\includegraphics[width=1 \textwidth]{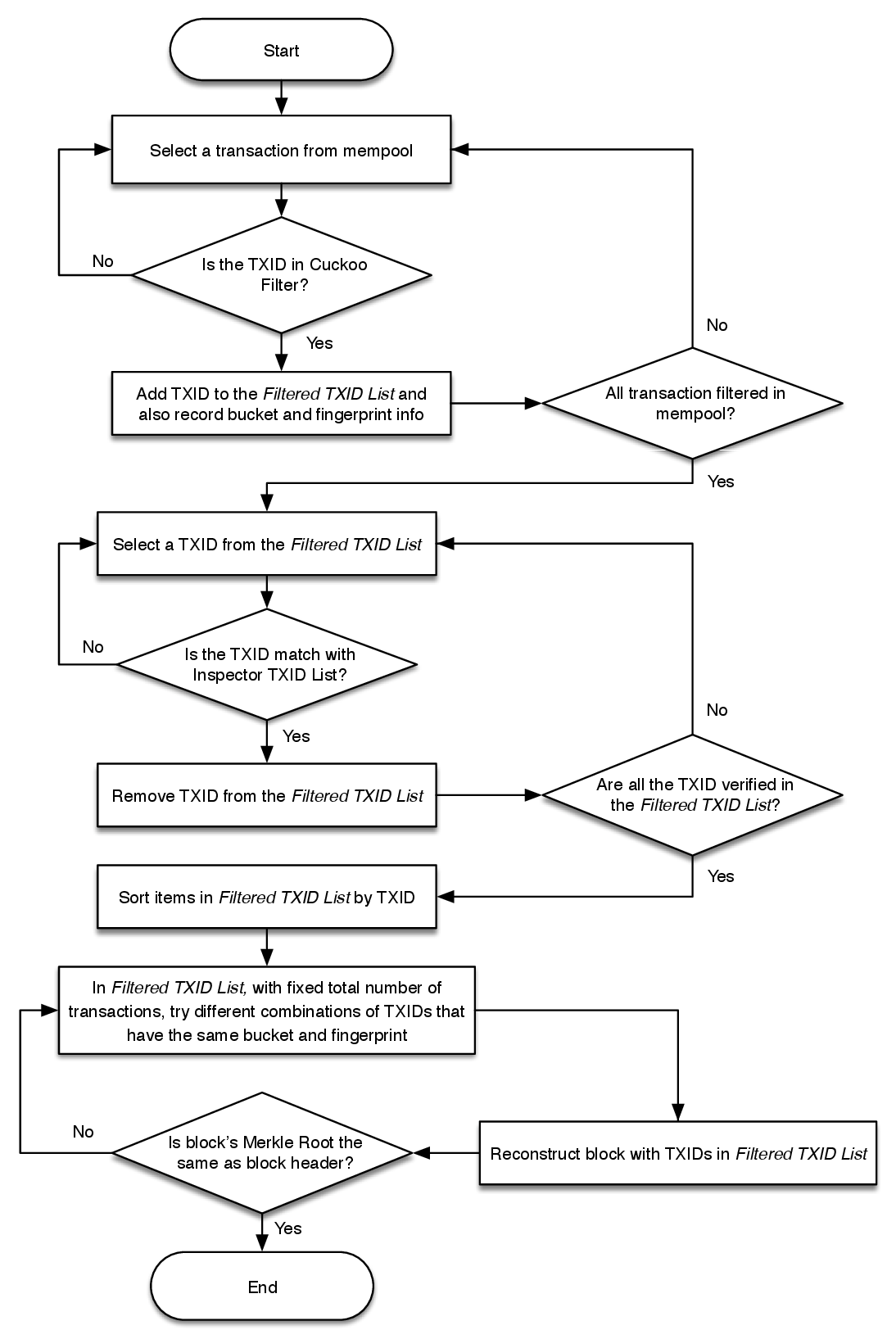}
		\caption{Flow chart of Receiver's mining process in the EDTS mechanism}
		\label{CompressionProcessing}
	\end{figure}
	
	\begin{algorithm}[htp] 
		\footnotesize
		\setstretch{1.3}
		\caption{Receiver's Block Reconstruct Process Under EDTS Mechanism} 
		\SetAlgoLined  
		\KwData{Mempool $M$, Efficient block $B$, Cuckoo Filter $C$, Transaction $T$, Filtered TXID List $FTList$, Fingerprint dictionary $Dict$ with $key$ and $value$, Number of transactions whose bucket and fingerprint only appear once in Cuckoo Filter $k$, Set of transactions whose bucket and fingerprint appear more than once in Cuckoo Filter $M$ and size $m$}
		\KwResult{Reconstruct block $RB$}
		\BlankLine 
		$k \leftarrow  0$\;
		$m \leftarrow  0$\;
		\ForEach{$T$ in $M$}{
			\If{$T \in C$}
			{
				Add $T$ to $FTList$\;
			}
		}
		\ForEach{$T$ in $FTList$}{
			\If{$T \in B$ is ``Inspector Transaction''}
			{
				Remove $T$ from $FTList$\;
			}
		}
		Sort $FTList$ by TXID\;
		\ForEach{$T$ in $FTList$}{
		    $key \leftarrow $T's bucket and fingerprint\;
		    $value \leftarrow $Count T's bucket and fingerprint\;
		    \ForEach{$key$ in $Dict$}{
    		    \eIf{$value$=1}
    		    {
                    $k \leftarrow k + 1$\;
                }{
                    $m \leftarrow m + value$\;
    		      Find $T$ with the $key$ and add to $M$
    		    }
            }
		}
		$j \leftarrow C^{N-k}_m$\;
		\For{$j > 0$}{
		    Select $m-(N-k)$ transactions from $M$\;
			Remove selected transactions from $FTList$\;
			Reconstruct block $RB$ using $FTList$\;
			\If{$RB.MerkelRoot = B.MerkelRoot$}
			{
				return $RB$\;
			}
			$j \leftarrow j - 1$\;
		}
	\end{algorithm}
	\restoregeometry
	
	\section{Experiments}
	\textcolor{black}{We have conducted a series of extensive simulation experiments to evaluate the proposed EDTS mechanism. Our experiments focus on two tasks:}

    \begin{itemize}

        \item \textcolor{black}{What is the optimal throughput that a DTS strategy can achieve under EDTS mechanism within the range of historical mining reward volatility?}

        \item \textcolor{black}{What are the optimal DTS strategy attribute settings for achieving the maximum throughput which can be adopted directly by the miners in the Bitcoin community without further optimization.}

    \end{itemize}

    These experiments are detailed in the following sections.

	\subsection{Experiment Settings}
	
	\textcolor{black}{The experiments were conducted on Huawei Cloud Stack (HCS) Ubuntu Linux v18.04 virtual machine with 48 core 2.5 GHz CPU, 1TB of ROM, and 192GB of RAM. In our experiments, SimBlock \cite{8751431} was used to evaluate the effectiveness of EDTS mechanism to improve throughput while maintaining low incentive volatility. Blocks in SimBlock are generated based on the Proof-of-Work consensus protocol and propagated on the simulated Blockchain network. SimBlock also allows the users to simulate other public Blockchain mining mechanism schemes, such as Proof-of-Burn (PoB), Proof-of-Stake (PoS), etc.}
	
	\textcolor{black}{In accordance with the current Bitcoin status, we list the parameter settings for our experiment in Table~\ref{SimBlockSettings}. We use the Bitcoin historic market data from December 2019 to September 2020. The historical data contains 400,000 transactions with minute-to-minute updates of OHLC (Open, High, Low, Close), transaction amount in BTC, and weighted Bitcoin price \cite{BitcoinHistoricalData}. Since the block size in Bitcoin is limited to 1MB, we also set the upper limit of the block size in EDTS mechanism to 1 MB. On average, each transaction can contain approximately 500 bytes of data. In our experiments, we adopt Bitcoin's one-year average number of transactions per block (i.e., approximately 2,100 transactions).}
	
	\linespread{1}
	\begin{table}[htp]
		\centering
		\footnotesize
		\caption{\textcolor{black}{Parameter settings for SimBlock Simulator}}
		\begin{threeparttable}
			\begin{tabular}{p{7.5cm}<{\raggedright} | p{5cm}<{\raggedright}}
				\toprule
				\textbf{Parameter} & \textbf{Value} \\
				\specialrule{0em}{3pt}{1pt}
				\hline  
				\specialrule{0em}{3pt}{1pt} 
				Block Size (byte)  & 1050000, 2100000, 4200000 \\
				\specialrule{0em}{3pt}{1pt}
				\hline  
				\specialrule{0em}{3pt}{1pt}
				Transaction Size (byte)  & 500 \\
				\specialrule{0em}{3pt}{1pt}
				\hline  
				\specialrule{0em}{3pt}{1pt} 
				New Transaction Rate & 3.5 Transactions per Second \\
				\specialrule{0em}{3pt}{1pt}
				\hline  
				\specialrule{0em}{3pt}{1pt} 
				Cuckoo Filter's false positive rate $\epsilon$ & $10^{-6}$ \\
				\specialrule{0em}{3pt}{1pt}
				\hline  
				\specialrule{0em}{3pt}{1pt} 
				Cuckoo Filter's hash table load factor $\alpha$ & $0.955$ \\
				\specialrule{0em}{3pt}{1pt}
				\hline  
			\end{tabular}
		\end{threeparttable}
		\label{SimBlockSettings}
	\end{table}
	
	In our experiments, we initializes the Blockchain with Bitcoin historical market transaction data and \textcolor{black}{add the data to} the mempool (a data structure that stores transactions waiting to be serialized). All nodes in the Bitcoin network share the same set of independent transactions that can be serialized in any order. These transactions are of the same size and they are incorporated into each blocks every 10 minutes. 
	
	\subsection{Attributes Considered in the Experiments} \label{Attributes}
	
	Attributes considered for designing EDTS mechanism are listed in Table~\ref{AttributeTable}. These attributes were previously reported in \cite{9592512}. \textcolor{black}{In the experiments, ``A2: Transaction Incorporation Priority'' and ``A4: Designated Space for Small-fee Transactions'' are discrete variables. Since other attributes are continuous variables, we divided the experiment into four parts according to the possible combination of A2 and A4. For ``A2: Transaction Incorporation Priority'', it indicates whether the priority of transaction incorporation was based on their arrival sequence or based on their transaction fee amount. For ``A4: Designated Space for Small-fee Transactions'', it indicates whether block space is reserved for high-priority transactions with small fees or not.}
	
	\linespread{1}
	\begin{table}[htp]
		\footnotesize
		\begin{center}
			\caption{Experiment settings for different DTS strategies attributes }
			\label{AttributeTable}
			\begin{tabular} {c  p{0.8cm}<{\centering}  p{2.5cm}<{\centering}  p{0.8cm}<{\centering}  p{0.8cm}<{\centering}  p{0.8cm}<{\centering}  p{0.8cm}<{\centering}  p{0.8cm}<{\centering}  p{0.8cm}<{\centering}  p{0.8cm}<{\centering} }
				\toprule
				\textbf{Exp.} & \textbf{A1} & \textbf{A2} & \textbf{A3} & \textbf{A4} & \textbf{A5} & \textbf{A6} & \textbf{A7} & \textbf{A8} & \textbf{A9} \\ [1.5ex] 
				\midrule
				1 & $\ast$ & Time-based & $\ast$  & False & N/A & N/A  & $\ast$ & $\ast$ & $\ast$ \\
				\specialrule{0em}{3pt}{1pt}
				\hline  
				\specialrule{0em}{3pt}{1pt}
				2 & $\ast$ & Time-based & $\ast$ & True & $\ast$ & $\ast$ & $\ast$ & $\ast$ & $\ast$ \\
				\specialrule{0em}{3pt}{1pt}
				\hline  
				\specialrule{0em}{3pt}{1pt}
				3 & $\ast$ & Fee-based & $\ast$ & False & N/A & N/A & $\ast$ & $\ast$ & $\ast$ \\
				\specialrule{0em}{3pt}{1pt}
				\hline  
				\specialrule{0em}{3pt}{1pt}
				4 & $\ast$ & Fee-based & $\ast$ & True & $\ast$ & $\ast$ & $\ast$ & $\ast$ & $\ast$ \\
				\specialrule{0em}{3pt}{1pt}
				\hline  
			\end{tabular}
			\begin{tablenotes}
				\item $\ast$ \textcolor{black}{Continuous variables;}\\A1: Mempool Size; A2: Transaction Incorporation Priority; A3: Transaction Fee Percentage; A4: Designated Space for Small-fee Transactions; A5: Small-fee Transaction Fee Threshold; A6: Small-fee Transaction Number Threshold; A7: Maximum Space for One Transaction; A8: Scale; A9: Shape; 
			\end{tablenotes}
		\end{center}
	\end{table}
	
	\subsection{Benchmarks}
	
	\textcolor{black}{In this paper, Throughput and Block Size as used as benchmarks for evaluating the scalability and storage requirement.} 
	
	\begin{itemize}
		\item \textbf{\textsl{Throughput.}} In the experiments, we measure the  ``Throughput" to evaluate the effectiveness of transactions confirmed/processed per second (TPS) by Bitcoin. The higher throughput, the wider the adoption of Bitcoin as a viable payment alternative. Because Bitcoin generates a 1MB block in 10 minutes interval, users typically have to wait six blocks to get the final state of the transaction. As a result, with typical transaction size, Bitcoin generates only about 3.5 transactions per second, far less throughput compare to VISA or Paypal.
		
		\item \textbf{\textsl{Block Size.}} Block size, relative to throughput, is another important metric for our experiment. In the Bitcoin network, blocks are created periodically, and each block contains a list of transactions. The number of transactions is limited by the block size, which affects throughput. However, as the block size increases, this block's probability of winning decreases, because smaller blocks tend to reach consensus more quickly. Nicolas Houy \cite{houy2014economics} also suggests that one cannot set arbitrarily large block sizes because it could jeopardize the very existence of Bitcoin in the long term. 
	\end{itemize}
	
	\subsection{Multi-objective Optimization}
	
	\textcolor{black}{ In \cite{9592512}, the authors proposed a set of Dynamic Transaction Storage (DTS) strategies. In the proposed EDTS mechanism, DTS strategies proposed in \cite{9592512} are combined with the Cuckoo Filter based storage mechanism proposed in the previous sections. However, not every DTS strategy is appropriate for using with EDTS mechanism. Therefore, in the following experiments, we adopt a Multi-objective Optimization algorithm to find the best strategies for integrating with the proposed EDTS mechanism. Note that the optimized strategies found in this experiment can be used/adopted by the miners in the Bitcoin community without further optimization. Specifically, real-time optimization by the miners is unnecessary when EDTS mechanism is used since the optimized strategies are generic and can be adopted by any miner. Therefore, there will be no additional overhead in computation for the miners in EDTS mechanism.}

	\textcolor{black}{In this paper, we focus on two objectives; volatility and throughput of EDTS mechanism. To achieve these objectives, Unified Non-dominated Sorting Genetic Algorithm \uppercase\expandafter{\romannumeral3} (U-NSGA-\uppercase\expandafter{\romannumeral3})~\cite{seada2015u} is adopted to obtain the Pareto front. U-NSGA-\uppercase\expandafter{\romannumeral3} was adopted in our experiments because of its superior performance in Pareto optimal searching and simplicity in parameters setting. However, any other state-of-art algorithms such as MOEA/D can be used to search for the Pareto front.}
	
	U-NSGA-\uppercase\expandafter{\romannumeral3} is an optimization algorithm for multi-objective optimizations based on one of the most popular algorithms, NSGA-\uppercase\expandafter{\romannumeral3}~\cite{deb2013evolutionary}. It is capable to solve mono-objective, bi-objective, and multi-objective optimization problems. Besides, niching-based tournament selection was used in U-NSGA-\uppercase\expandafter{\romannumeral3} to improve its performance for mono-objective and bi-objective optimizations while maintaining the same ability in higher dimensional optimization~\cite{seada2015u}. The detailed process of U-NSGA-\uppercase\expandafter{\romannumeral3} is described in~\ref{NSGA}.
	
	As mentioned in section~\ref{Attributes}, there are 9 attributes in the EDTS mechanism which can influence the two objectives considered in this paper namely the volatility and the throughput. A combination of 9 attributes is regarded as the position of a point to be optimized in U-NSGA-\uppercase\expandafter{\romannumeral3}. SimBlock is a Java program but the optimization package Pymoo \cite{pymoo} that we adopted is in written in Python. Therefore the U-NSGA-\uppercase\expandafter{\romannumeral3}-based Optimizer (UNO) was implemented in Python to link the simulator with Pymoo. The summarized process of UNO is as follows:
	
	\begin{enumerate}
		\item \textsl{\textbf{Initialization:}} U-NSGA-\uppercase\expandafter{\romannumeral3} initializes a population of $N$ points with different positions. Each position of a point represents a different combination of the 9 EDTS attributes.
		\item \textsl{\textbf{Execution:}} UNO passes the obtained positions as arguments to run the SimBlock $N$ times.
		\item \textsl{\textbf{Simulation:}} By using the inputs from UNO, SimBlock starts to simulate the the behavior of the miners and then returns the results of the block incentives.
		\item \textsl{\textbf{Evaluation:}} UNO monitors the simulations and retrieves the result of each point to compute the volatility and the throughput. Based on the collected volatility and throughput of $N$ points, U-NSGA-\uppercase\expandafter{\romannumeral3} computes the current non-dominant front and creates the next generation of points.
	\end{enumerate}
	
	Due to a large number of points and generations needed to be processed, the experiment requires long computational time to complete an entire round of optimization. To alleviate this problem, in the experiments, a multiprocessing design similar to the PSO-based optimizer is adopted in UNO. Specifically, in one generation, $n$ simulations are executed in parallel until all $N$ positions are exhausted. 
	
	\textcolor{black}{The s-Energy method proposed in~\cite{blank2020generating} adopts \refeqs{s-energy} where $s$ is the dimension of a many-dimensional space to generate the reference directions in the same size for U-NSGA-\uppercase\expandafter{\romannumeral3}. Therefore the algorithm requires only two parameters, the size of the population and a termination criterion. A popular choice of population is between 100 and 200. Considering the optimization is in a nine dimensional space, a population size of 100 is considered to be adequate. In our preliminary testing, the results usually converge in less than 100 generations. Therefore, the maximum generation is set to 100.} 
	
	\begin{equation}
		U(z^{(i)}, z^{(j)}) = \frac{1}{||z^{(i)}-z^{(j)}||^s} \label{s-energy}
	\end{equation}
	
	\subsection{Experiment Results}

    In the experiments, we search the optimal combination of attributes in a continuous space to find the highest TPS and lowest volatility. Table~\ref{OptimizationResult} lists the U-NSGA-\uppercase\expandafter{\romannumeral3} multi-objective optimization results of EDTS mechanism. In the corresponding figures for each experiment (see the rightmost column of Table~\ref{OptimizationResult}), we show the correlation between TPS and volatility under different DTS attribute settings. \textcolor{black}{In these figures, historical volatility range which is depicted in light blue color shading which is obtained from historical volatility range (see Table~\ref{HistoricalVolatility} in Section 4). Note that the lower and upper bound of the range are calculated based on Bitcoin’s average daily block rewards from 2012 to 2021.}
	
	\linespread{1}
	\begin{table}[htp]
		\footnotesize
		\begin{center}
			\caption{TPS and volatility correlations for four optimization experiments}
			\label{OptimizationResult}
			\begin{tabular}{ p{1cm}<{\centering}| p{4cm}<{\centering}  |p{4cm}<{\centering}| p{3cm}<{\centering}}
				\toprule%
				\textbf{Exp.}  & \textbf{TPS} & \textbf{Volatility} & \textbf{Result Figure} \\ [1ex] 
				\midrule
				\textbf{1} & \textbf{\textcolor{black}{135.0 - 325.3} TPS} & \textbf{0.050589 - 0.235770} & \textbf{\reffig{Exp1}} \\
				\specialrule{0em}{3pt}{1pt}
				\hline  
				\specialrule{0em}{3pt}{1pt}
				2 & \textcolor{black}{86.1 - 322.7} TPS & 0.042374 - 0.237429 & \reffig{Exp2} \\
				\specialrule{0em}{3pt}{1pt}
				\hline  
				\specialrule{0em}{3pt}{1pt}
				3 & \textcolor{black}{40.8 - 237.3} TPS & 0.047383 - 0.234076 & \reffig{Exp3}\\
				\specialrule{0em}{3pt}{1pt}
				\hline  
				\specialrule{0em}{3pt}{1pt}
				4 & \textcolor{black}{42.0 - 264.6} TPS & 0.047509 - 0.237046  & \reffig{Exp4} \\
				\specialrule{0em}{3pt}{1pt}
				\hline  
			\end{tabular}
		\end{center}
	\end{table}
	
	\newgeometry{left = 3.6 cm, top=0.8cm}
	\begin{figure}[htp]       
		\subfigure 
		{
			\begin{minipage}{5cm}
				\centering          
				\includegraphics[scale=0.56]{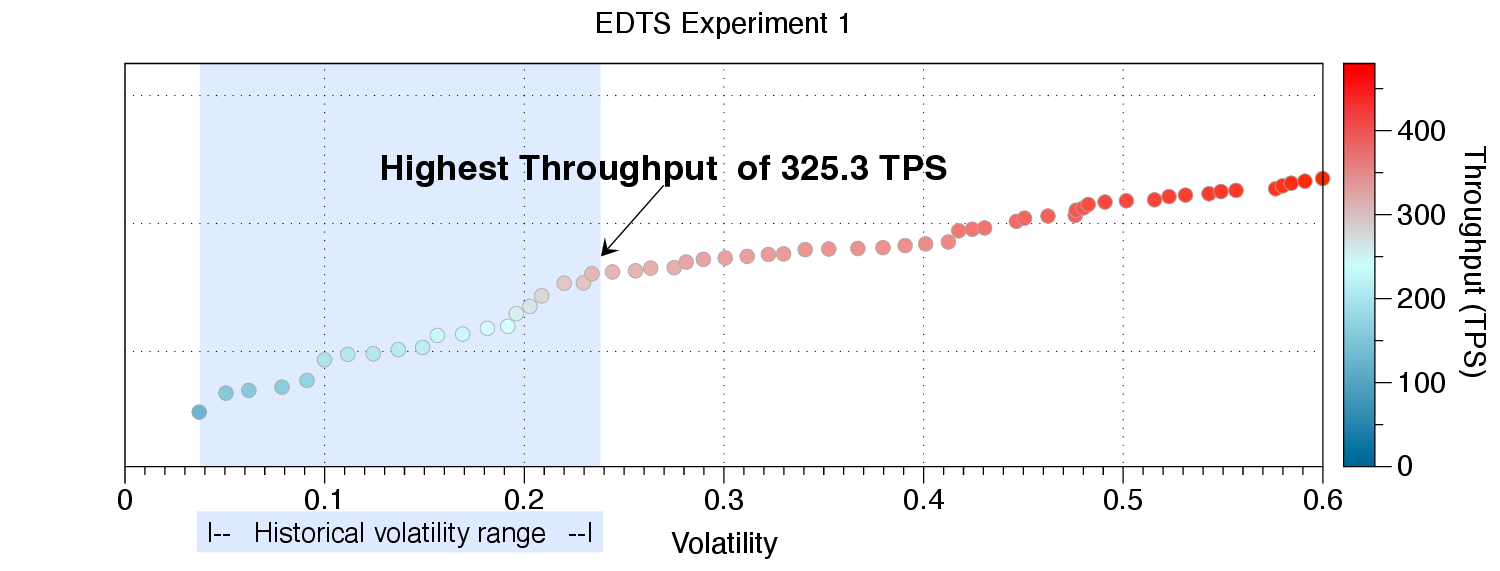}   
			\end{minipage}
		}
		\subfigure 
		{
			\begin{minipage}{8cm}
				\centering      
				\includegraphics[scale=0.01]{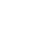}   
			\end{minipage}
		} 
		\subfigure 
		{
			\begin{minipage}{5cm}
				\centering          
				\includegraphics[scale=0.56]{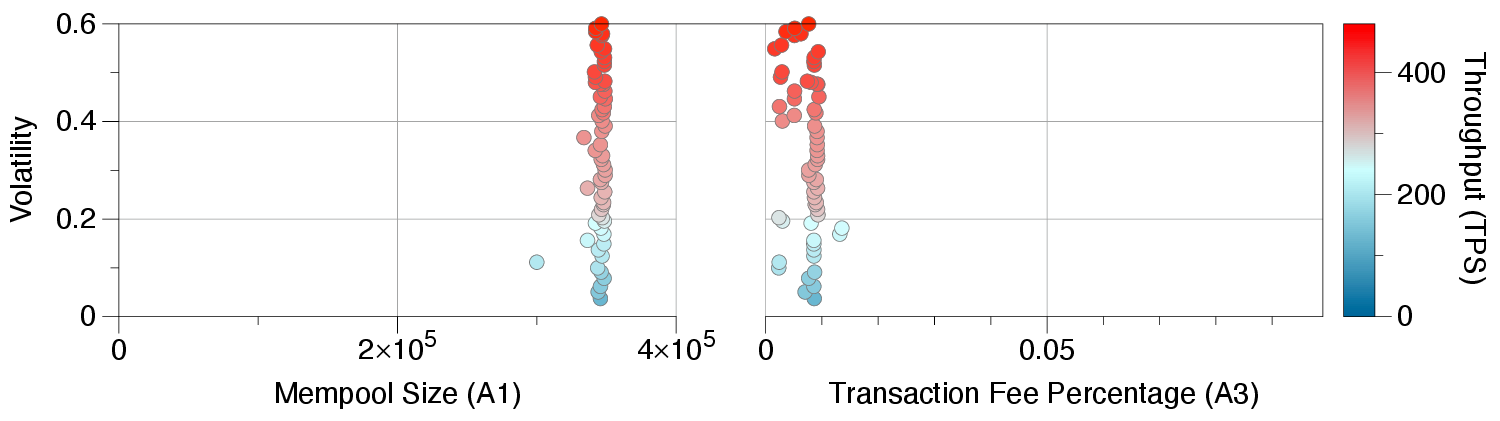}   
			\end{minipage}
		}
		\subfigure 
		{
			\begin{minipage}{8cm}
				\centering      
				\includegraphics[scale=0.01]{blank.eps}   
			\end{minipage}
		} 
		\subfigure 
		{
			\begin{minipage}{5cm}
				\centering          
				\includegraphics[scale=0.56]{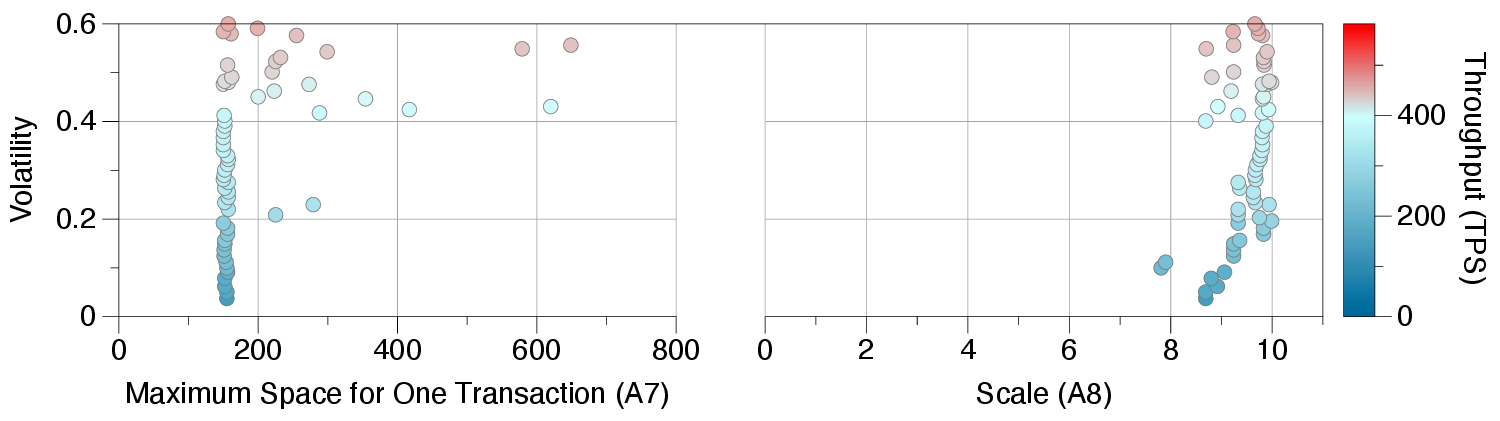}   
			\end{minipage}
		}
		\subfigure 
		{
			\begin{minipage}{8cm}
				\centering      
				\includegraphics[scale=0.01]{blank.eps}   
			\end{minipage}
		} 
		\subfigure 
		{
			\begin{minipage}{5cm}
				\centering          
				\includegraphics[scale=0.56]{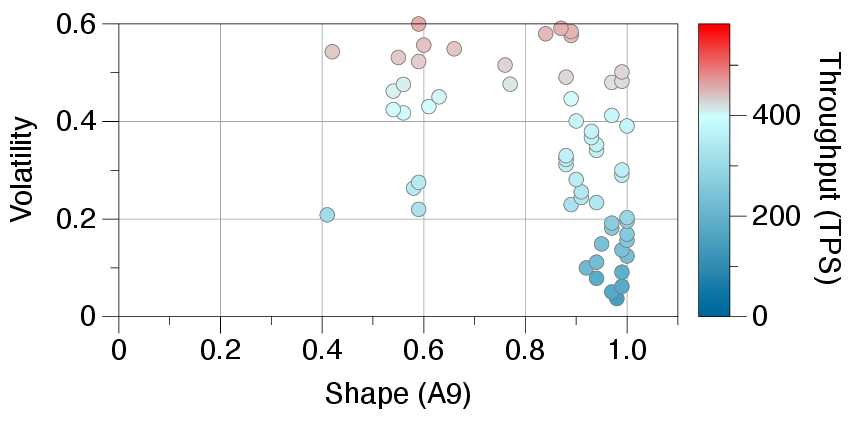}   
			\end{minipage}
		}
		\subfigure 
		{
			\begin{minipage}{8cm}
				\centering      
				\includegraphics[scale=0.01]{blank.eps}   
			\end{minipage}
		} 
		\subfigure 
		{
			\begin{minipage}{5cm}
				\centering          
				\includegraphics[scale=1]{blank.eps}   
			\end{minipage}
		}
		\subfigure 
		{
			\begin{minipage}{8cm}
				\centering      
				\includegraphics[scale=0.01]{blank.eps}   
			\end{minipage}
		}
		\caption{The throughput and volatility optimization results of Experiment 1 and the correlation between different attribute values and throughput in the same volatility range} 
		\label{Exp1}  
	\end{figure}
	
	\begin{figure}[htp]       
		\subfigure 
		{
			\begin{minipage}{5cm}
				\centering          
				\includegraphics[scale=0.56]{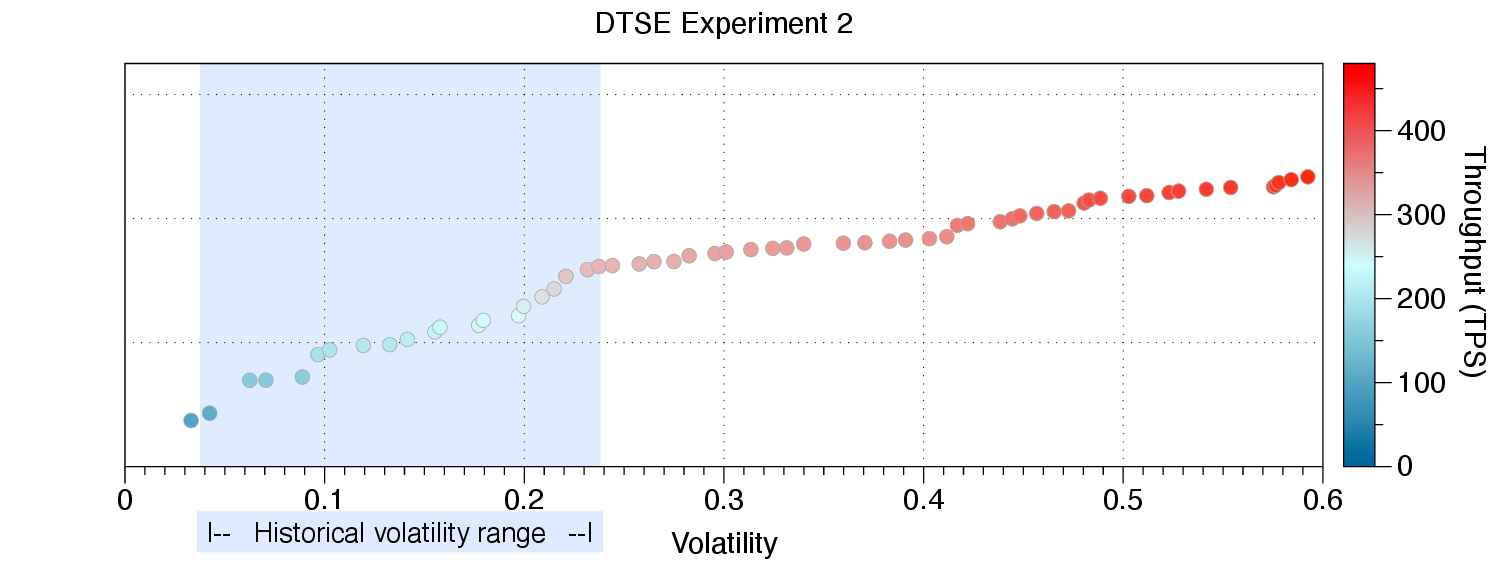}   
			\end{minipage}
		}
		\subfigure 
		{
			\begin{minipage}{8cm}
				\centering      
				\includegraphics[scale=0.01]{blank.eps}   
			\end{minipage}
		} 
		\subfigure 
		{
			\begin{minipage}{5cm}
				\centering          
				\includegraphics[scale=0.56]{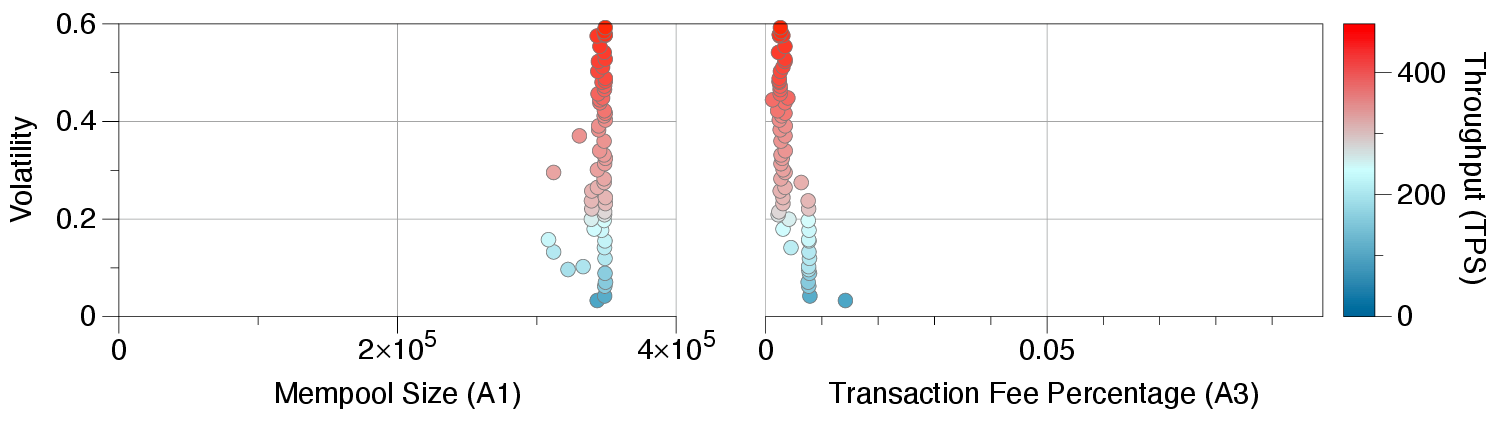}   
			\end{minipage}
		}
		\subfigure 
		{
			\begin{minipage}{8cm}
				\centering      
				\includegraphics[scale=0.01]{blank.eps}   
			\end{minipage}
		} 
		\subfigure 
		{
			\begin{minipage}{5cm}
				\centering          
				\includegraphics[scale=0.56]{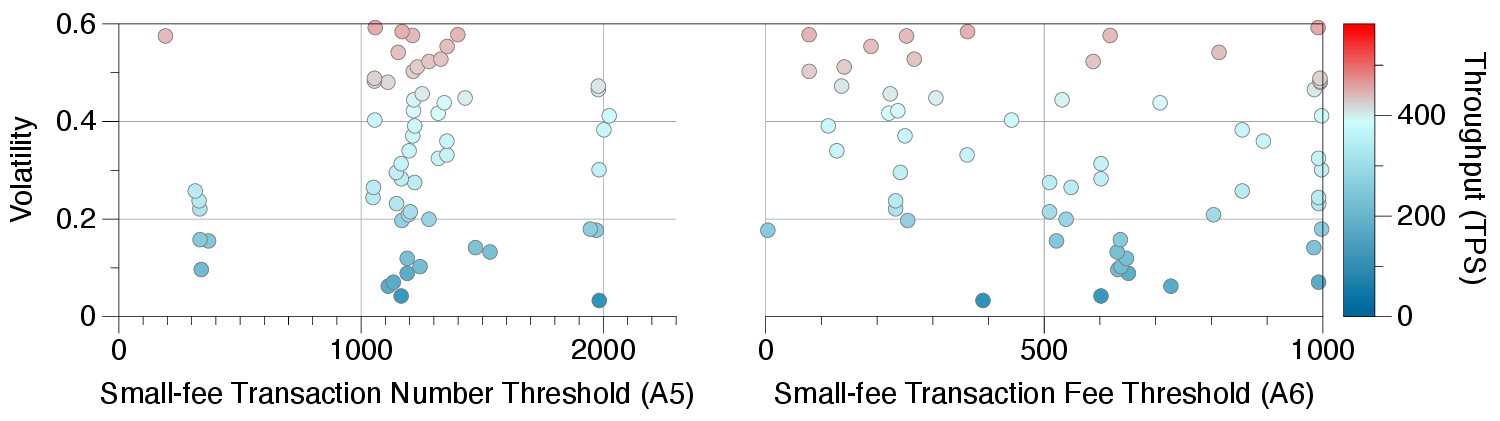}   
			\end{minipage}
		}
		\subfigure 
		{
			\begin{minipage}{8cm}
				\centering      
				\includegraphics[scale=0.01]{blank.eps}   
			\end{minipage}
		} 
		\subfigure 
		{
			\begin{minipage}{5cm}
				\centering          
				\includegraphics[scale=0.56]{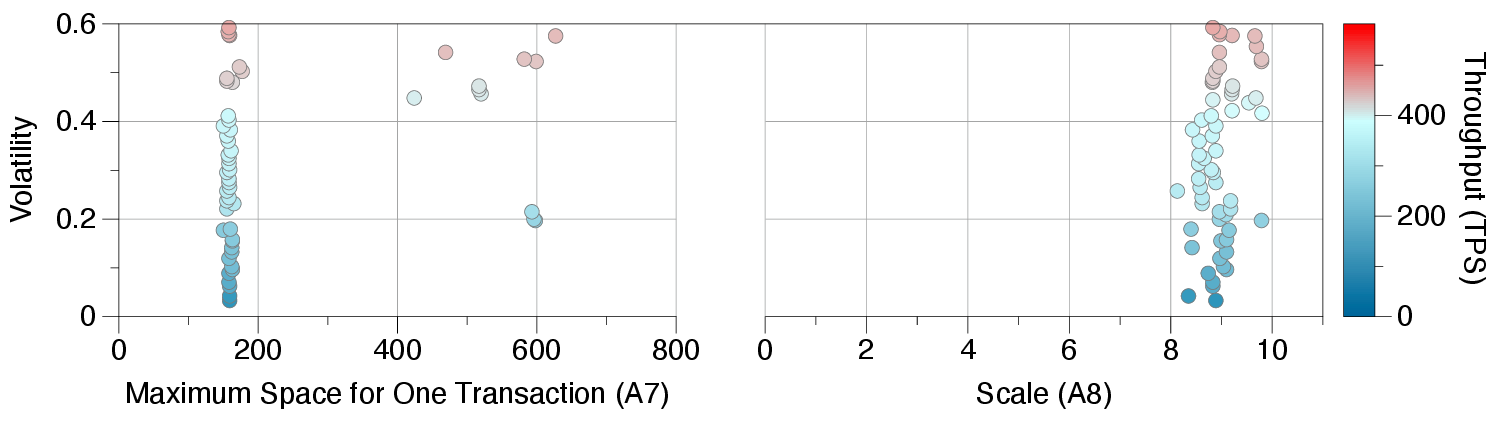}   
			\end{minipage}
		}
		\subfigure 
		{
			\begin{minipage}{8cm}
				\centering      
				\includegraphics[scale=0.01]{blank.eps}   
			\end{minipage}
		} 
		\subfigure 
		{
			\begin{minipage}{5cm}
				\centering          
				\includegraphics[scale=0.56]{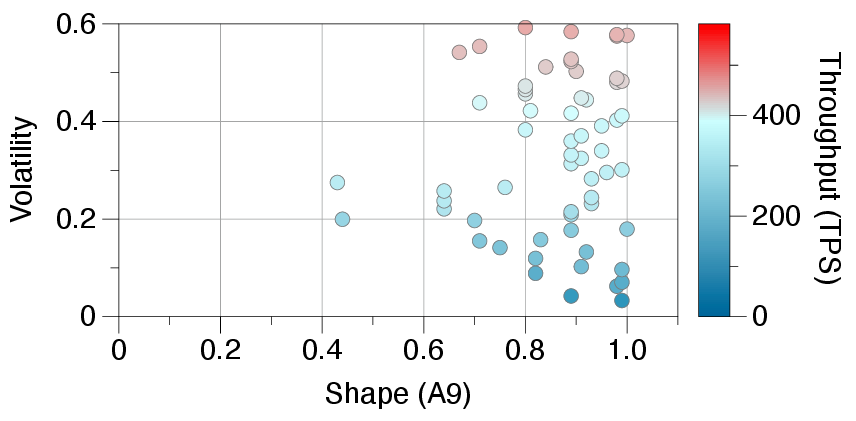}   
			\end{minipage}
		}
		\subfigure 
		{
			\begin{minipage}{8cm}
				\centering      
				\includegraphics[scale=0.01]{blank.eps}   
			\end{minipage}
		} 
		\caption{The throughput and volatility optimization results of Experiment 2 and the correlation between different attribute values and throughput in the same volatility range} 
		\label{Exp2}  
	\end{figure}
	
	\begin{figure}[htp]
		\subfigure 
		{
			\begin{minipage}{5cm}
				\centering          
				\includegraphics[scale=0.56]{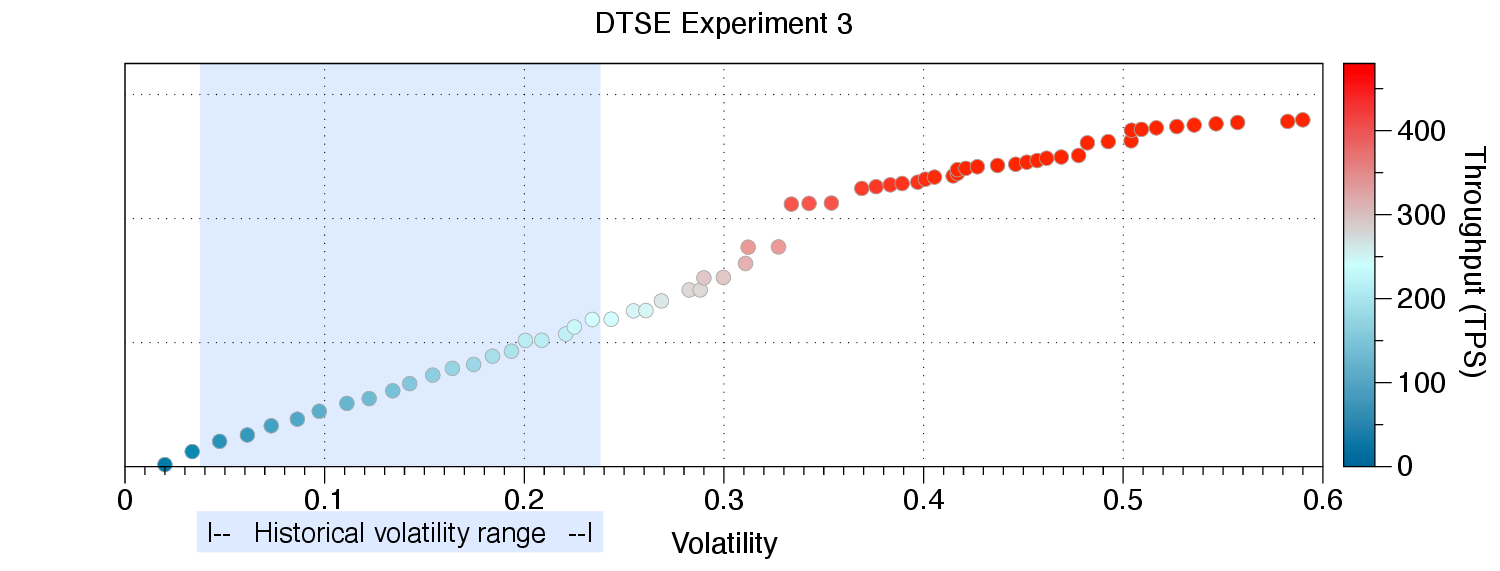}   
			\end{minipage}
		}
		\subfigure 
		{
			\begin{minipage}{8cm}
				\centering      
				\includegraphics[scale=0.01]{blank.eps}   
			\end{minipage}
		} 
		\subfigure 
		{
			\begin{minipage}{5cm}
				\centering          
				\includegraphics[scale=0.56]{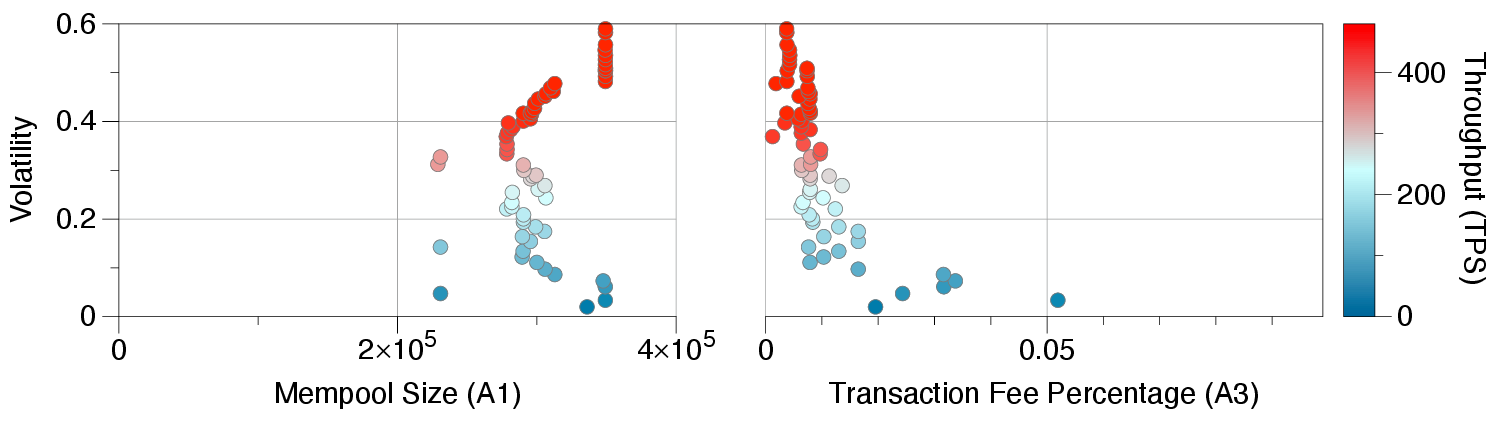}   
			\end{minipage}
		}
		\subfigure 
		{
			\begin{minipage}{8cm}
				\centering      
				\includegraphics[scale=0.01]{blank.eps}   
			\end{minipage}
		} 
		\subfigure 
		{
			\begin{minipage}{5cm}
				\centering          
				\includegraphics[scale=0.56]{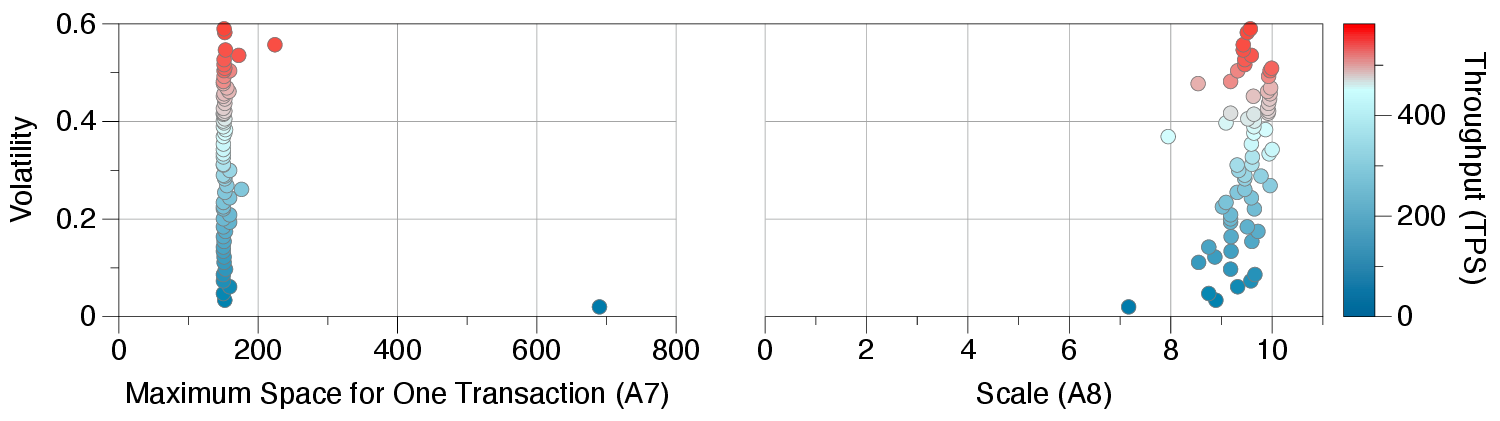}   
			\end{minipage}
		}
		\subfigure 
		{
			\begin{minipage}{8cm}
				\centering      
				\includegraphics[scale=0.01]{blank.eps}   
			\end{minipage}
		} 
		\subfigure 
		{
			\begin{minipage}{5cm}
				\centering          
				\includegraphics[scale=0.56]{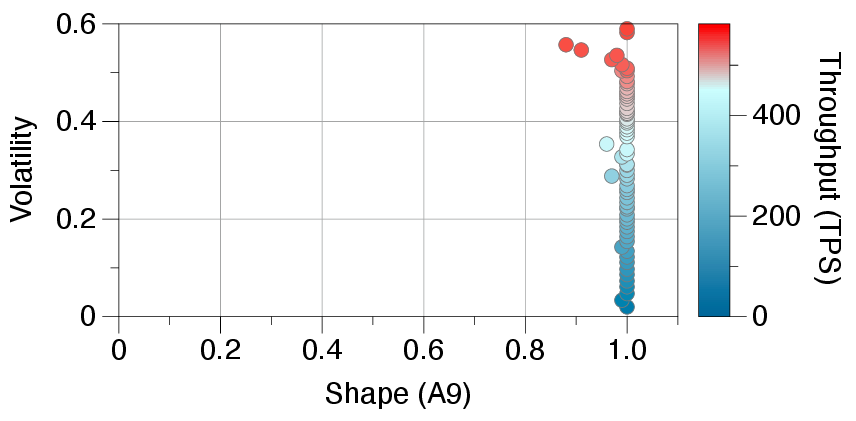}   
			\end{minipage}
		}
		\subfigure 
		{
			\begin{minipage}{8cm}
				\centering      
				\includegraphics[scale=0.01]{blank.eps}   
			\end{minipage}
		}  
		\subfigure 
		{
			\begin{minipage}{5cm}
				\centering          
				\includegraphics[scale=1]{blank.eps}   
			\end{minipage}
		}
		\subfigure 
		{
			\begin{minipage}{8cm}
				\centering      
				\includegraphics[scale=0.01]{blank.eps}   
			\end{minipage}
		}
		\caption{The throughput and volatility optimization results of Experiment 3 and the correlation between different attribute values and throughput in the same volatility range} 
		\label{Exp3}  
	\end{figure}
	
	\begin{figure}[htp]       
		\subfigure 
		{
			\begin{minipage}{5cm}
				\centering          
				\includegraphics[scale=0.56]{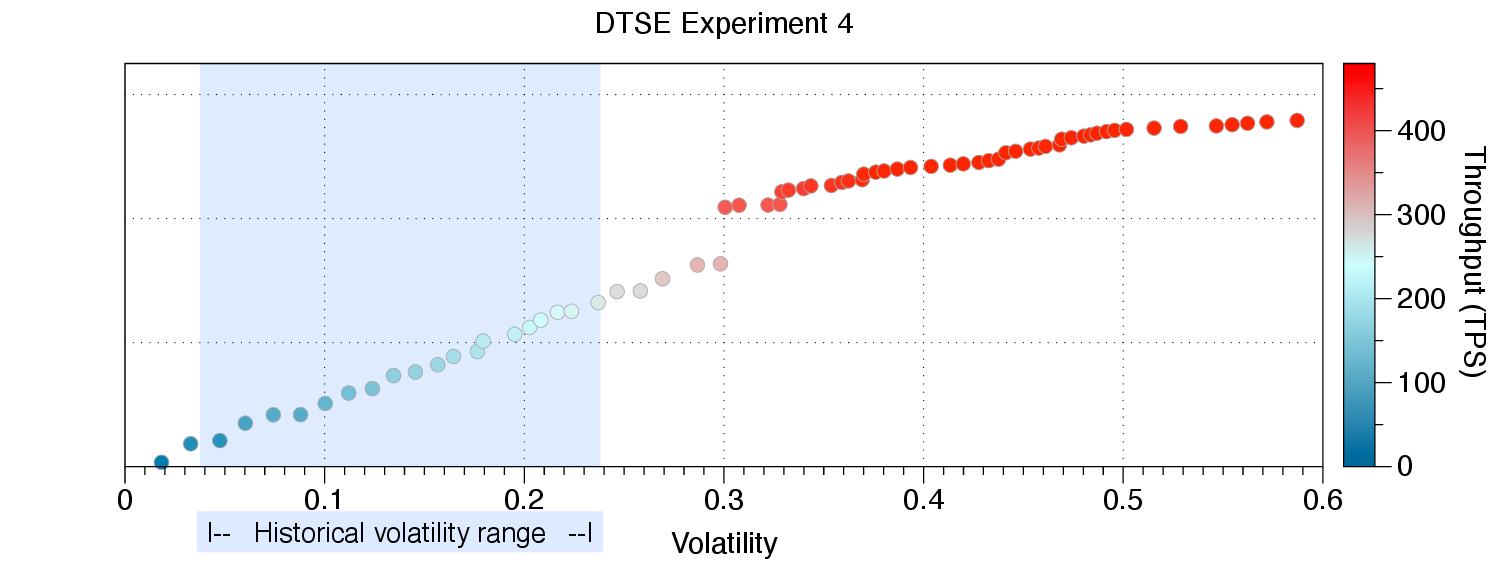}   
			\end{minipage}
		}
		\subfigure 
		{
			\begin{minipage}{8cm}
				\centering      
				\includegraphics[scale=0.01]{blank.eps}   
			\end{minipage}
		} 
		\subfigure 
		{
			\begin{minipage}{5cm}
				\centering          
				\includegraphics[scale=0.56]{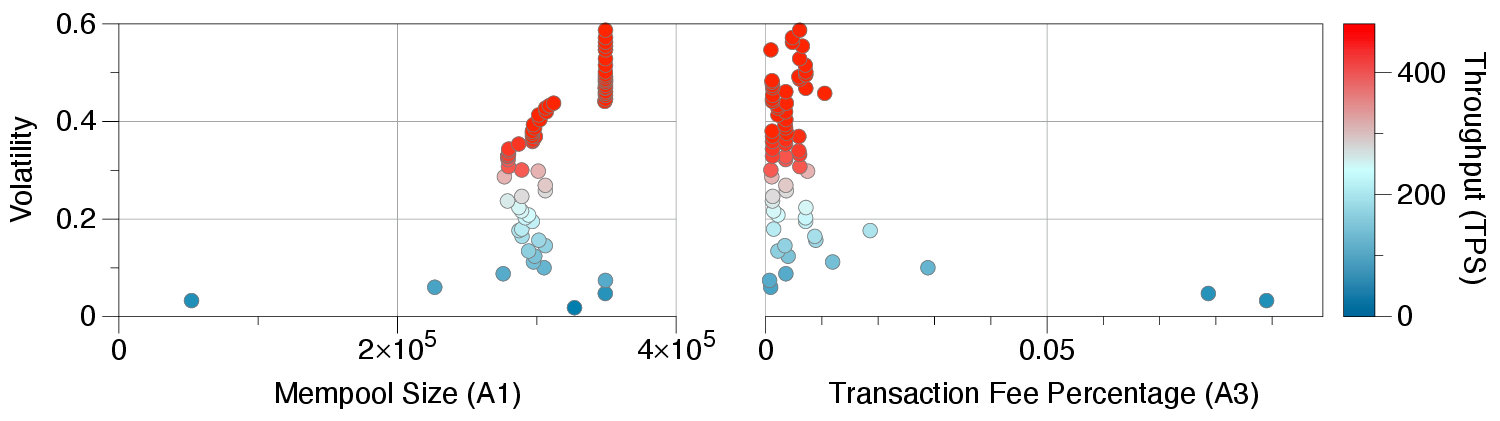}   
			\end{minipage}
		}
		\subfigure 
		{
			\begin{minipage}{8cm}
				\centering      
				\includegraphics[scale=0.01]{blank.eps}   
			\end{minipage}
		} 
		\subfigure 
		{
			\begin{minipage}{5cm}
				\centering          
				\includegraphics[scale=0.56]{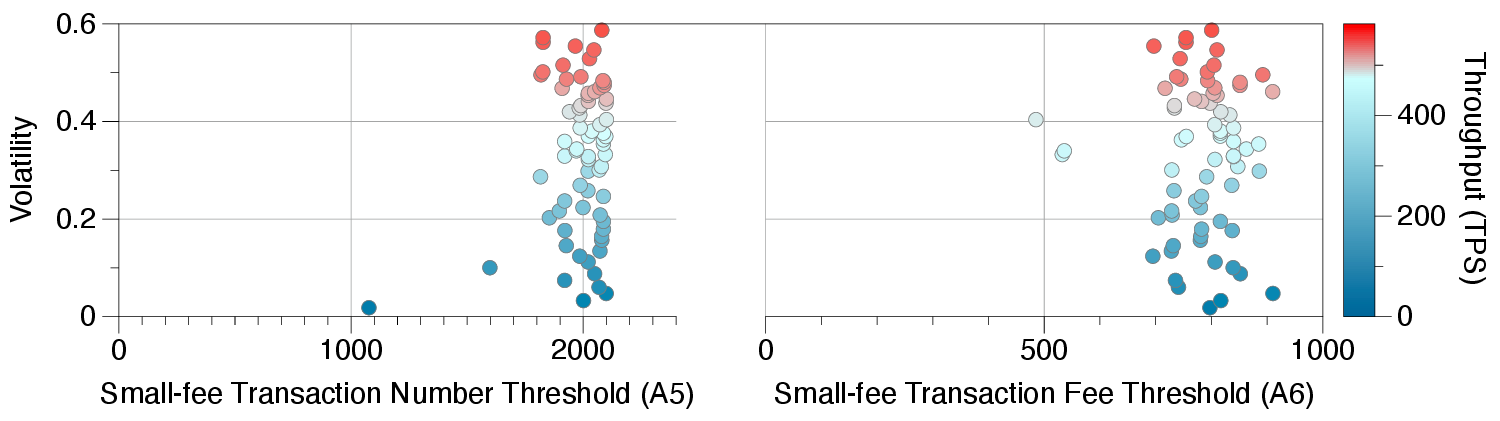}   
			\end{minipage}
		}
		\subfigure 
		{
			\begin{minipage}{8cm}
				\centering      
				\includegraphics[scale=0.01]{blank.eps}   
			\end{minipage}
		} 
		\subfigure 
		{
			\begin{minipage}{5cm}
				\centering          
				\includegraphics[scale=0.56]{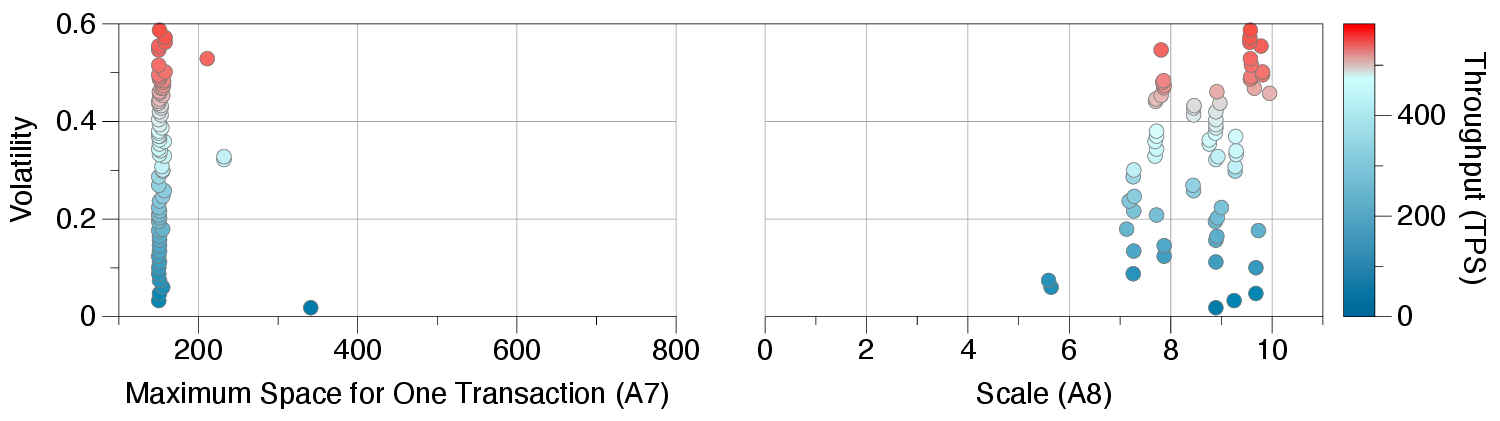}   
			\end{minipage}
		}
		\subfigure 
		{
			\begin{minipage}{8cm}
				\centering      
				\includegraphics[scale=0.01]{blank.eps}   
			\end{minipage}
		} 
		\subfigure 
		{
			\begin{minipage}{5cm}
				\centering          
				\includegraphics[scale=0.56]{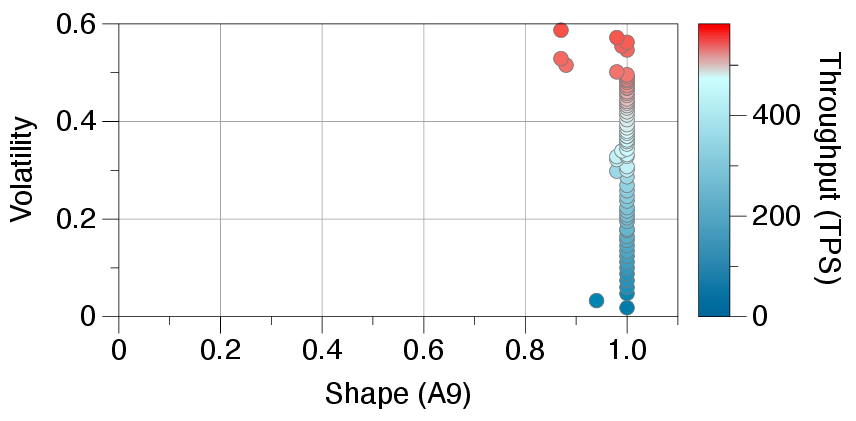}   
			\end{minipage}
		}
		\subfigure 
		{
			\begin{minipage}{8cm}
				\centering      
				\includegraphics[scale=0.01]{blank.eps}   
			\end{minipage}
		} 
		\caption{The throughput and volatility optimization results of Experiment 4 and the correlation between different attribute values and throughput in the same volatility range} 
		\label{Exp4}  
	\end{figure}
	\restoregeometry
	
	\textcolor{black}{Probabilistic method such as Cuckoo Filter significantly reduces the space cost of an item. According to Fan et al. \cite{fan2014cuckoo}, the upper bound of space cost (bytes) the Cuckoo Filter required is irrelevant to the size of the input transaction and it can be calculated by \refeqs{cuckoo_cost}. A transaction in a Cuckoo Filter with $\epsilon=10^{-6}$ and $\alpha=0.955$ can occupy a maximum storage space of 3.0015 Byte, which is a significant space saving compared to the average size of the original transaction (500 bytes).}
	
	\textcolor{black}{From Table~\ref{HistoricalVolatility} (see Section 4), we can observe that the historical mining reward volatility are within the range of 0.037647-0.238111. When the block reward volatility fluctuates within the historical range, the throughput of Experiment 1 was better than other three experiments, reaching 135.0-325.3 TPS. By referencing the historical mining volatility range, the optimization results obtained in Expirement 1 in \reffig{Exp1} show that the maximum throughput of EDTS is approximately 325.3 TPS. All the DTS strategy attribute settings are list in Table~\ref{Exp1AttributeTable} when throughput reach 325.3 TPS in Expirement 1. According to the setting in Experiment 1, transactions are incorporated in the block according to their arrival sequence, and there is no reserved block space for high-priority transactions with small fees. These results reveal that EDTS is the able to achieve high throughput and efficient block propagation while keeping the Bitcoin sustainable.}

    \linespread{1}
    \begin{table}[htp]
        \begin{center}
            \footnotesize
            \caption{\textcolor{black}{DTS strategy attributes setting when throughput reaches 325.3 TPS in Experiment 1 (\reffig{Exp1})}}
            \label{Exp1AttributeTable}
            \begin{tabular} { p{1.2cm}<{\centering} p{1.5cm}<{\centering}  p{1.2cm}<{\centering}  p{1cm}<{\centering}  p{1cm}<{\centering}  p{1cm}<{\centering}  p{1cm}<{\centering}  p{1cm}<{\centering}  p{1cm}<{\centering} }
                \toprule
                \textbf{A1} & \textbf{A2} & \textbf{A3} & \textbf{A4} & \textbf{A5} & \textbf{A6} & \textbf{A7} & \textbf{A8} & \textbf{A9} \\ [1.5ex] 
                \midrule
                75032 & Fee-based & 10.31\% & False & N/A & N/A & 36 & 9.5 & 0.99 \\
                \specialrule{0em}{3pt}{1pt}
                \hline  
            \end{tabular}
            \begin{tablenotes}
                \item A1: Mempool Size; A2: Transaction Incorporation Priority; A3: Transaction Fee Percentage; A4: Designated Space for Small-fee Transactions; A5: Small-fee Transaction Fee Threshold; A6: Small-fee Transaction Number Threshold; A7: Maximum Space for One Transaction; A8: Scale; A9: Shape; 
            \end{tablenotes}
        \end{center}
    \end{table}
	
	\section{Comparison with on-chain scaling solutions }
	
	This section is devoted for comparing the proposed EDTS mechanism with the Bitcoin scalability solutions introduced in Section\uppercase\expandafter{\romannumeral2}. Despite the fact that existing solutions are making progress in improving the scalability, solutions introduced in Section\uppercase\expandafter{\romannumeral2} have been proposed under the regime where the coin-based block reward is still dominant. Therefore, existing solutions will need to be re-examined in terms of sustainability as new challenges emerge. In this section, we analyze the proposed EDTS with other Bitcoin scalability solutions from following perspectives: Scalability, Sustainability and Prioritization. In Table~\ref{EvaluationTable}, we list the comparison results in terms of each baseline category, including: throughput, block size and latency. 
	
	\linespread{0.9}
	\begin{table}[htp]
		\footnotesize
		\centering
		\caption{\textcolor{black}{On-chain scaling solutions comparison}}
		\label{EvaluationTable}
		\begin{tabular}{p{3cm}<{\centering} |p{4cm}<{\centering} |p{3cm}<{\centering}  }
			\toprule%
			\textbf{Solution} & \textbf{Throughput (TPS)} & \textbf{Block Size (MB)} \\ [1ex]
			\midrule
			MAST & 3.5 & -- \\
			\specialrule{0em}{3pt}{1pt}
			\hline
			\specialrule{0em}{3pt}{1pt} 
			Segwit & 7 & 4 \\
			\specialrule{0em}{3pt}{1pt}
			\hline
			\specialrule{0em}{3pt}{1pt}  
			Bitcoin Cash & 61 & 32 \\
			\specialrule{0em}{3pt}{1pt}  
			\hline
			\specialrule{0em}{3pt}{1pt} 
			Bitcoin NG & 100 & NA \\
			\specialrule{0em}{3pt}{1pt}
			\hline
			\specialrule{0em}{3pt}{1pt}  
			EDTS & \textcolor{black}{135.0 - 325.3} & $<1^\ast$ \\
			\specialrule{0em}{3pt}{1pt}
			\hline  
		\end{tabular}
		\begin{tablenotes}
			\item NA - Not Applicable, ``-" - indicates Not available.\\
			$\ast$ The parameters change dynamically based on the number of transactions incorporated in the block and the urgency (fee) of each transaction.\\
		\end{tablenotes}
	\end{table}
	
	\subsection{Scalability}
	
	In recent years, a number of on-chain scaling solutions were proposed. These solutions mainly focus on increasing the number of transactions per block. On-chain solutions such as MAST, Segwit, Bitcoin Cash, and Bitcoin NG attempt to achieve faster throughput by modifying the overall consensus or changing the way transaction data is stored. 
	
	However, existing solutions have their limitations. For example, in MAST, throughput improvement is not enough to solve the scalability problem whereas in Segwit and Bitcoin Cash, scalability is improved by incorporating more transactions into a block. Nevertheless, as the block size increases, propagation speed becomes slow, increasing the probability of orphan blocks and maintenance costs, resulting in the centralization of mining. Although Bitcoin NG can handle transactions of up to 100 TPS, it faces the threat that the incentives of micro-blocks are incompatible with the investment of miners.
	
	The proposed EDTS mechanism uses Cuckoo Filters to generate space-efficient block, limiting the amount of data that needs to be propagate across the network. When the maximum block size is limited to 1MB, as our experimental results in Table~\ref{OptimizationResult} show, the EDTS mechanism can achieve a throughput of \textcolor{black}{325.3} TPS (\reffig{Exp1}) when the volatility of EDTS is within the Bitcoin historical volatility range. \textcolor{black}{Due to the Cuckoo Filter's efficiency in compression, even if more transactions are added to a block in EDTS, the block size grows much more slowly than Segwit and Bitcoin Cash. Moreover, EDTS can handle more than three times transactions that of Bitcoin NG without mismatched mining investment problem.}
	
	\subsection{Sustainability}
	
	Since Bitcoin block reward is halved every four years, transaction fees are expected to become the main source of miners' revenue. Carlsten et al. \cite{10.1145/2976749.2978408} reported several issues when transaction fees dominate mining rewards. Therefore, it is crucial to investigate whether mining reward mechanism can prevent deviant mining strategies from threatening the integrity and security of the Blockchain.
	
	After coin-based rewards reach zero, the transaction-fee regime can lead to potential instability in Bitcoin's block reward. Specifically, the time-varying nature of transaction fees allows a richer set of strategic deviations. To ensure that security is commensurate with increased transaction volume, Bitcoin networks need to provide miners with sustainable incentives in terms of transaction fees to keep them engaged and competitive. By stabilizing mining incentives for each block under the transaction-fee regime, DTS strategies~\cite{9592512} could eliminate strategic deviations such as \textsl{Selfish Mining}, \textsl{Undercutting}, \textsl{Mining Gap}, and \textsl{Pool Hopping}.
	
	Existing solutions such as MAST, Segwit, and Bitcoin Cash either change the transaction storage mechanism or adopt larger block sizes to increase the throughput. Besides, solutions such as Bitcoin NG, Lightning Network, still rely on Bitcion's mining rewards to incentivize mining activities for leadership elections or to register final settlement transactions on the chain. However, these solution have separate channels to process transactions. For sharding solutions such as Elastico, mining is still conducted in different shards. All of the above activities are based on the coin-based reward currently available on Bitcoin. Therefore, when the Bitcoin mining transitions to a transaction-fee regime, the integrity of Bitcoin networks that adopt existing solutions will be threatened by deviant Mining behavior.
	
	Whereas in our posed EDTS mechanism, the block space that can be occupied by a transaction can be dynamically calculated according to transaction cost and parameter setting of DTS mechanism. Each transaction is then added to the Cuckoo Filter and the size of the final result of the Cuckoo Filter varies depending on the number of transactions that can be incorporated into a block. By utilizing the DTS strategies, the volatility of the block reward remains stable. Therefore, EDTS mechanism can improve the Bitcoin's throughput while stabilizing the block incentives and preventing deviant miner behavior. In summary, the proposed EDTS mechanism enables the Bitcoin mining to be sustainable under the transaction-fee regime.
	
	\subsection{Prioritization}
	
	Existing scalability solutions often focus on scaling the Bitcoin with main emphasis on the consensus. However, network-level block propagation is also considered as an essential step in the transaction synchronization process. After the new block is mined, it will be propagated across the Bitcoin network using a ``peer-to-peer propagation mechanism" which is an ``information diffusion process" inherent in the Blockchain. A miner sends the blocks to 8-10 peers, whom further forward them to their peers until the blocks are synchronized across the whole network. \textcolor{black}{In the current Bitcoin network, the link bandwidth for blocks to reach each node through the gossip protocol determines the travel time of the blocks in the network.}
	
	Transaction fees play an important role in prioritizing transactions. Users typically submit transactions with a certain amount of fees to get their desired priority. 
	The proposed EDTS mechanism is consistent with existing solutions when \textsl{Transaction Incorporation Priority} attribute is set to \textsl{Fee-based}. In this setting, miners first select and pack those transactions with higher priorities as their mining basis. In contrast to EDTS, the existing mechanism is limited to improving the consensus mechanism and their block size is uniform. Under the existing mechanism, even if the overall transaction fees of blocks are different, the propagation process of blocks has no differentiated priority. As for EDTS, large transactions which offer higher transaction fees are not only preferentially incorporated in the block at the mining stage, they can also reach the consensus stage faster during the block propagation process. 
	
	Due to the limitation in the Bitcoin network's bandwidth, the propagation time of a block is proportional to the size of the block. Therefore, block propagation time $P_n$ from one node to another can also vary significantly. Since the nodes's coordinates are randomly determined through $Reg(N)$. Block propagation time from node $N_{f}$ to node $N_{t}$ can be formulated as follows:
	
	\begin{normalsize}
		\begin{equation}
			\resizebox{.85\hsize}{!}{$P_n =  \displaystyle\frac{Size(B)}{Bandwidth(Reg(N_{f}),Reg(N_{t}))} + Delay(Reg(N_{f}),Reg(N_{t}))$}
			\label{propagation}
		\end{equation}
	\end{normalsize}
	
	\textcolor{black}{For a block to be propagated through the entire Blockchain, it needs to be propagated $M$ times. We denote the total propagation time for a block as $PT$} (see \refeqs{total_propagation}).
	
	\begin{normalsize}
		\begin{equation}
			\resizebox{.15\hsize}{!}{$PT=\displaystyle\sum_{n=1}^MP_n$}
			\label{total_propagation}
		\end{equation}
	\end{normalsize}
	
\begin{figure}[htp]
		\makeatletter
		\def\@captype{figure}
		\makeatother
		\centering
		\includegraphics[width=1 \textwidth]{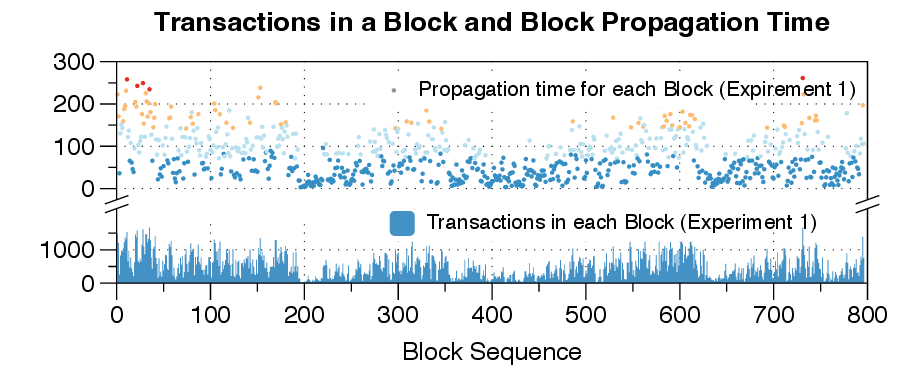}
		\caption{Under the EDTS mechanism, the number of transactions incorporated in different blocks, as well as the difference in the propagation time of corresponding blocks}
		\label{PropagationTime}
	\end{figure}

	\textcolor{black}{To better illustrate the results, we plot the block propagation time of different blocks based on Expirement 1 (see \reffig{Exp1}). Other Experiments can also be plotted in the same way.} In \reffig{PropagationTime}, the lower part of the figure represents the number of transactions incorporated into a block under the EDTS mechanism, while the upper part of the figure represents how much time different block takes to propagate though the whole Bitcoin network. From \reffig{PropagationTime}, we can observe that higher-fee transactions will be incorporated into smaller blocks, and smaller blocks tend to reach consensus faster. Therefore EDTS can provide more efficient P2P transaction relays for higher fee transactions by taking advantage of the dynamic block size feature. This also suggests that proposed EDTS mechanism can stimulate investors' willingness to commit higher transaction fees.
	
	\section{Conclusion}
	
	In this article, we propose a novel Efficient Dynamic Transaction Storage (EDTS) mechanism for improving the throughput in the Bitcoin network based on the optimized Dynamic Transaction Storage (DTS) strategies. In the proposed mechanism, Cuckoo Filter is used to minimize the transaction storage. We also adopt U-NSGA-\uppercase\expandafter{\romannumeral3} multi-objective optimization algorithm to find the optimal attributes for the Dynamic Transaction Storage (DTS) strategies. Unlike existing solutions, EDTS mechanism was designed to address both volatility and throughput objectives simultaneously. To evaluate the proposed mechanism, extensive experiments were conducted based on the Bitcoin historical data. Experimental results reveal that the proposed mechanism can achieve the high throughput while maintaining the volatility within the historical range, which is considered to be the stable. 
	Besides, we further elaborate the advantages of the EDTS mechanism from three perspectives; Scalability, Sustainability, and Prioritization. 
	
	In summary, the proposed EDTS mechanism has a unique advantage in keeping Bitcoin sustainable. It is also immune to deviant mining behavior when Bitcoin transits into the transaction-fee regime. When the maximum block size is limited to 1MB like in Bitcoin, the EDTS mechanism can achieve a throughput of \textcolor{black}{325.3} TPS (with respect to the historical volatility range). Moreover, to the best of authors' knowledge, EDTS is the first scaling solution that enables the Bitcoin to be sustainable while increasing the throughput when Bitcoin transits to the transaction-fee regime in the future. As for the future work, we are planning to integrating EDTS with off-chain scaling solutions such as Lightning Network~\cite{poon2015Bitcoin} or Sharding solutions such as Elastico~\cite{luu2016secure} to further improve the scalability of Bitcoin.
	
	\section*{Acknowledgment}
	This research was funded by University of Macau (File no. MYRG2019-00136-FST).

	\pagebreak
	\appendix
	
	\section{U-NSGA-\uppercase\expandafter{\romannumeral3}}\label{NSGA}
	
	A Pareto optimal is a situation where there is no single objective can be better off without making the other objective worse off, and a Pareto front consists of a set of such optimal points. In our context, trade-offs between lower volatility and higher throughput should be decided according to the Pareto front given by an evolutionary multi-objective optimization (EMO) algorithm. The evolution process of U-NSGA-\uppercase\expandafter{\romannumeral3} is as follows.
	
	\begin{enumerate}
		\item $N$ points are selected randomly covering the $D$-dimensional searching space and a set of reference points $H$ having a normal vector of one on a hyperplane is generated. For $p$ divisions and therefore $p+1$ directions on a single objective, the size of $H$ for $M$ objectives is $\tbinom{M+p-1}{p}$. The reference points are placed by the s-Energy method proposed by J. Blank et. all~\cite{blank2020generating}, which can generate an arbitrary number of points.
		
		\item In the current generation, the set of reserve parents is $P_t$ with size $N$, from which points are selected into the mating pool $P_t^{'}$ by two rounds of niching-based tournament selection. The procedure traverses the reserve parents $P_t$ to hold a tournament for every two points and finally select $N/2$ points who have the right to mate. In one tournament, if two points are not associated with the same reference directions, one of them is selected randomly. Otherwise, the point from the higher nondomination level is selected. Besides, in the situation where the two points are from the same niche and nondomination level, the one with a shorter distance to the reference directions is selected. After shuffling and traversing the set of reserve parents one more time, remaining $N/2$ points are selected into the mating pool $P_t^{'}$. 
		
		\item Parents from the mating pool $P_t^{'}$ are chosen to create their offspring $Q_t$ with the size of $N$ by recombination and mutation. The reserved parents $P_t$ are combined with the offspring $Q_t$ to form $R_t$ with a size of $2N$, in which all the points are sorted according to their nondomination levels (i.e. $F_1$ for the non-dominated level, $F_2$ for the second level, and so on). 
		
		\item Points from the first level $F_1$ to the final level $F_l$ are selected into $S_t$ until the size of $S_t$ is equal to or bigger than $N$. Next, all points from the level bigger than $F_l$ are excluded from $S_t$. When $\lvert S_t\rvert =N$, the algorithm proceeds to the next generation with $P_{t+1}$=$S_t$. However, it is more often that $\lvert S_t\rvert > N$. In this case, only a part of the points in $F_l$ are selected to $P_{t+1}$. Step 5 to 7 describes the reference-directions-based niching that assists in selecting the rest of $K$ points from $F_l$.
		
		\item At the beginning, $P_{t+1}$ is the combination of $F_1$ to $F_{l-1}$, and the rest is $K=N-\lvert P_{t+1}\rvert$. the objectives of all the points in $S_t$ are first normalized. From the current population $S_t$, the ideal point $z_t^{min}$ is identified as $\Bar{z^{min}}=(z^{min}_1, z^{min}_2, \cdots, z^{min}_M)$ for all objectives from the first to $M$th. The points with the maximum value in $j$th objective is identified as extreme point $z^{j,max}$ from which the intercept $a_j$ of $j$th objective can be calculated. All objectives $f_j$ from 1 to M are then normalized using \refeqs{normalize} to form a hyperplane. Because the reference directions generated by s-Energy method are already on the normalized plane, there is no more normalization for them.
		\begin{equation}
			f_j^n(x)=\frac{f_j-z_i^{min}}{a_i} \label{normalize}
		\end{equation}
		\item After normalization, for each point in the population $S_t$, the perpendicular distance from the point to each reference direction is computed, and the closest reference direction is associated with the point. The distance $d^{\perp}(s,w)$ for a point $s$ to a reference direction $w$ is given by \refeqs{perpendicular}.
		\begin{equation}
			d^{\perp}(s,w)=\lVert (s-w^{T}sw/\lVert w\rVert^2)\rVert \label{perpendicular}
		\end{equation}
		\item For $j$th reference direction $\Bar{j}$, the number of points from $P_{t+1}$ (i.e. points from the last level $F_l$ which are not in $P_{t+1}$) are counted as niche count $\rho_j$. If there is no point from $P_{t+1}$ that is associated with the reference direction $\Bar{j}$ but one or more points from $F_l$ are associated with $\Bar{j}$, the point from $F_l$ and closest to the direction $\Bar{j}$ is selected, and the niche count $\rho_j$ is increased by 1. A reference direction is out of consideration for the present generation if neither a point from $P_{t+1}$ nor from $F_l$ is associated with this direction. Otherwise, if $\rho_j\geq1$, a point that is from $F_l$ and associated with the reference direction $\Bar{j}$ is randomly selected, and the niche count $\rho_j$ of this direction is incremented by 1. The previous procedure is iterated $K$ times until $K$ members are selected and added to $P_{t+1}$.
		\item Step 1 to 7 are iterated until the maximum generation is reached. 
	\end{enumerate}

	
		
		
		

\begin{thebibliography}{10}
\expandafter\ifx\csname url\endcsname\relax
  \def\url#1{\texttt{#1}}\fi
\expandafter\ifx\csname urlprefix\endcsname\relax\def\urlprefix{URL }\fi
\expandafter\ifx\csname href\endcsname\relax
  \def\href#1#2{#2} \def\path#1{#1}\fi

\bibitem{nakamoto2008Bitcoin}
S.~Nakamoto, {Bitcoin: A Peer-to-Peer Electronic Cash System} (2008) 9.

\bibitem{ethereum.org}
Ethereum.org.
\newblock \href{https://ethereum.org/}{{Ethereum}} [online].
\newblock (Last accessed: 13 March 2022).

\bibitem{litecoin}
Coinsutra.com, {Litecoin}, \url{https://coinsutra.com/litecoin-cryptocurrency},
  Last accessed: 13 March 2022.

\bibitem{10.1145/2382196.2382292}
G.~O. Karame, E.~Androulaki, S.~Capkun, {Double-Spending Fast Payments in
  Bitcoin}, in: Proceedings of the 2012 ACM Conference on Computer and
  Communications Security, CCS '12, Association for Computing Machinery, New
  York, NY, USA, 2012, p. 906–917.

\bibitem{10.1145/2976749.2978408}
M.~Carlsten, H.~Kalodner, S.~M. Weinberg, A.~Narayanan, {On the Instability of
  Bitcoin Without the Block Reward}, in: Proceedings of the 2016 ACM SIGSAC
  Conference on Computer and Communications Security, CCS '16, Association for
  Computing Machinery, New York, NY, USA, 2016, p. 154–167.

\bibitem{eyal2016Bitcoin}
I.~Eyal, A.~E. Gencer, E.~G. Sirer, R.~Van~Renesse, $\{$Bitcoin-NG$\}$: A
  scalable blockchain protocol, in: 13th USENIX symposium on networked systems
  design and implementation (NSDI 16), 2016, pp. 45--59.

\bibitem{rubin2014merkelized}
J.~Rubin, M.~Naik, N.~Subramanian, Merkelized abstract syntax trees,
  XP055624837, Dec 16 (2014) 3.

\bibitem{lombrozo2015segregated}
E.~Lombrozo, J.~Lau, P.~Wuille, Segregated witness (consensus layer), Bitcoin
  Core Develop. Team, Tech. Rep. BIP 141 (2015).

\bibitem{poon2015Bitcoin}
J.~Poon, T.~Dryja, The bitcoin lightning network, Scalable o-chain instant
  payments (2015).

\bibitem{luu2016secure}
L.~Luu, V.~Narayanan, C.~Zheng, K.~Baweja, S.~Gilbert, P.~Saxena, A secure
  sharding protocol for open blockchains, in: Proceedings of the 2016 ACM
  SIGSAC Conference on Computer and Communications Security, 2016, pp. 17--30.

\bibitem{9592512}
X.~Zhao, Y.-W. Si, Dynamic transaction storage strategies for a sustainable
  blockchain, in: 2021 IEEE International Conference on Services Computing
  (SCC), 2021, pp. 309--318.

\bibitem{9133427}
A.~Hafid, A.~S. Hafid, M.~Samih, Scaling blockchains: A comprehensive survey,
  IEEE Access 8 (2020) 125244--125262.

\bibitem{seada2015u}
H.~Seada, K.~Deb, U-nsga-iii: a unified evolutionary optimization procedure for
  single, multiple, and many objectives: proof-of-principle results, in:
  International conference on evolutionary multi-criterion optimization,
  Springer, 2015, pp. 34--49.

\bibitem{basu2019stablefees}
S.~Basu, D.~Easley, M.~O'Hara, E.~Sirer, Stablefees: A predictable fee market
  for cryptocurrencie, Available at SSRN 3318327 (2019).

\bibitem{10.1145/3479722.3480991}
M.~V.~X. Ferreira, D.~J. Moroz, D.~C. Parkes, M.~Stern, Dynamic Posted-Price
  Mechanisms for the Blockchain Transaction-Fee Market, Association for
  Computing Machinery, New York, NY, USA, 2021, p. 86–99.

\bibitem{10.1145/3308558.3313454}
R.~Lavi, O.~Sattath, A.~Zohar, Redesigning bitcoin's fee market, in: The World
  Wide Web Conference, WWW '19, Association for Computing Machinery, New York,
  NY, USA, 2019, p. 2950–2956.

\bibitem{EIP1559}
R.~D. M. S. I. N. A.~B. Vitalik~Buterin, Eric~Conner, Ethereum improvement
  proposals, no. 1559, april 2019. [online serial],
  \url{https://eips.ethereum.org/EIPS/eip-1559}, Last accessed: 13 Jan 2022.

\bibitem{leonardos2021dynamical}
S.~Leonardos, B.~Monnot, D.~Reijsbergen, E.~Skoulakis, G.~Piliouras, Dynamical
  analysis of the eip-1559 ethereum fee market, in: Proceedings of the 3rd ACM
  Conference on Advances in Financial Technologies, 2021, pp. 114--126.

\bibitem{Bitcoincash}
{Bitcoincash.com}, {Bitcoin Cash}, \url{https://bitcoincash.org}, Last
  accessed: 20 Jan 2021.

\bibitem{10.1007/978-3-662-45472-5_28}
I.~Eyal, E.~G. Sirer, {Majority Is Not Enough: Bitcoin Mining Is Vulnerable},
  in: N.~Christin, R.~Safavi-Naini (Eds.), Financial Cryptography and Data
  Security, Springer Berlin Heidelberg, Berlin, Heidelberg, 2014, pp. 436--454.

\bibitem{DBLP:journals/corr/abs-1112-4980}
M.~Rosenfeld, {Analysis of Bitcoin Pooled Mining Reward Systems}, CoRR
  abs/1112.4980 (2012).

\bibitem{6688704}
C.~Decker, R.~Wattenhofer, Information propagation in the bitcoin network, in:
  IEEE P2P 2013 Proceedings, 2013, pp. 1--10.

\bibitem{fan2014cuckoo}
B.~Fan, D.~G. Andersen, M.~Kaminsky, M.~D. Mitzenmacher, Cuckoo filter:
  Practically better than bloom, in: Proceedings of the 10th ACM International
  on Conference on emerging Networking Experiments and Technologies, 2014, pp.
  75--88.

\bibitem{10.1145/3319535.3354237}
G.~Naumenko, G.~Maxwell, P.~Wuille, A.~Fedorova, I.~Beschastnikh, Erlay:
  Efficient transaction relay for bitcoin, in: Proceedings of the 2019 ACM
  SIGSAC Conference on Computer and Communications Security, CCS '19,
  Association for Computing Machinery, New York, NY, USA, 2019, p. 817–831.

\bibitem{nasdaq}
Nasdaq.com, {Total value of coinbase block rewards and transaction fees paid to
  miners.},
  \url{https://data.nasdaq.com/data/BCHAIN/MIREV-bitcoin-miners-revenue}, Last
  accessed: 13 Jan 2022.

\bibitem{btc}
Btc.com, {Bitcoin Blocks List}, \url{https://btc.com/btc/blocks}, Last
  accessed: 13 Jan 2022.

\bibitem{8751431}
R.~{Banno}, K.~{Shudo}, {Simulating A Blockchain Network with SimBlock}, in:
  2019 IEEE International Conference on Blockchain and Cryptocurrency (ICBC),
  2019, pp. 3--4.

\bibitem{BitcoinHistoricalData}
{Kaggle.com}, {Bitcoin Historical Dataset},
  \url{https://www.kaggle.com/datasets/prasoonkottarathil/btcinusd}, Last
  accessed: 20 Jan 2022.

\bibitem{houy2014economics}
N.~Houy, {The Economics of Bitcoin Transaction Fees}, GATE WP 1407 (2014).

\bibitem{deb2013evolutionary}
K.~Deb, H.~Jain, An evolutionary many-objective optimization algorithm using
  reference-point-based nondominated sorting approach, part i: solving problems
  with box constraints, IEEE transactions on evolutionary computation 18~(4)
  (2013) 577--601.

\bibitem{pymoo}
J.~{Blank}, K.~{Deb}, pymoo: Multi-objective optimization in python, IEEE
  Access 8 (2020) 89497--89509.

\bibitem{blank2020generating}
J.~Blank, K.~Deb, Y.~Dhebar, S.~Bandaru, H.~Seada, Generating well-spaced
  points on a unit simplex for evolutionary many-objective optimization, IEEE
  Transactions on Evolutionary Computation 25~(1) (2020) 48--60.

\end{thebibliography}
\end{document}